\documentclass[twocolumn, prd, aps, tightenlines, preprintnumbers, showpacs, nofootinbib, superscriptaddress, notitlepage, superscriptaddress]{revtex4-1}

\usepackage{pgfplots}
\usepackage{filecontents}
\usepackage{standalone}
\usepackage{graphicx}
\usepackage{amsmath,amsthm,amssymb,amsfonts}

\usepackage{pgfplotstable}
\usepackage{bm}
\usepackage{slashed}
\usepackage{epsfig}
\usepackage{amsfonts}
\usepackage{color}
\usepackage{epstopdf}
\usepackage[pdftex]{hyperref}

\newcommand{\non}{\nonumber\\}
\newcommand\mo{{\mathcal O}}
\newcommand{\state}[4]{{^#1\hspace{-0.6mm}#2_{#3}^{[#4]}}}

\newcommand\CScSa{\state{3}{S}{1}{1}}

\newcommand\COaSz{\state{1}{S}{0}{8}}

\newcommand\COcPz{\state{3}{P}{0}{8}}

\definecolor{mygreen1}{RGB}{0,73,44}
\definecolor{mygreen2}{RGB}{0,140,81}

\def\empile#1\above#2{\mathrel{\mathop{\kern 0pt#1}\limits_{#2}}}

\pgfplotsset{compat=1.10}

\begin{document}

%\preprint{Preprint \#}

\title{Event engineering studies for heavy flavor production and hadronization in high multiplicity hadron-hadron and hadron-nucleus collisions}
\date{\today}
\author{Yan-Qing Ma}
\affiliation{
School of Physics and State Key Laboratory of Nuclear Physics and
Technology, Peking University, Beijing 100871, China.}
\affiliation{
Center for High Energy Physics,
Peking University, Beijing 100871, China.}
\affiliation{Collaborative Innovation Center of Quantum Matter,
Beijing 100871, China.}
\author{Prithwish Tribedy}
\author{Raju Venugopalan}
\affiliation{
Physics Department, Brookhaven National Laboratory, Upton, New York 11973-5000, USA.}
\author{Kazuhiro Watanabe}
\affiliation{
Physics Department, Old Dominion University, Norfolk, Virginia 23529, USA.}
\affiliation{
Theory Center, Jefferson  Laboratory, Newport News, Virginia 23606, USA.}
\date{\today}

\begin{abstract}
Heavy flavor measurements in high multiplicity proton-proton and proton-nucleus collisions at collider energies enable unique insights into their production and hadronization mechanism because experimental and theoretical uncertainties cancel in ratios of  their cross-sections relative to minimum bias events. We explore such event engineering using the Color Glass Condensate (CGC) effective field theory to compute short distance charmonium cross-sections.  The CGC is combined with heavy-quark fragmentation functions to compute $D$-meson cross-sections; for the $J/\psi$, hadronization is described employing Nonrelativistic QCD (NRQCD) and an Improved Color Evaporation model. Excellent agreement is found between the CGC computations and the LHC heavy flavor data in high multiplicity events. Event engineering in this CGC+NRQCD framework reveals a very rapid growth in the fragmentation of the $^3S_1^{[8]}$ state in rare events relative to minimum bias events.

\end{abstract}
%11.80.La 	Multiple scattering
%12.38.Bx 	Perturbative calculations
%13.87.Fh 	Fragmentation into hadrons
%14.40.Lb	Charmed mesons (|C|>0, B=0)
%14.40.Pq 	Heavy quarkonia

\maketitle

%%%%%%%%%

%%%%%%%%%%%%%%%%%%%%%%%%%%%%%%%%%%%%%%%%%%%%%%%%%%%%%%%%%%%%%
\section{Introduction}{\label{section:introduction}}

The study of high multiplicity events in proton-proton ($p+p$) and proton-nucleus ($p+A$) collisions at the Relativistic Heavy Ion Collider (RHIC) and the Large Hadron Collider (LHC) has focused attention on the spatial and momentum structure of rare parton configurations in the colliding projectiles obtained by variations in the multiplicity, energy and system size. Such ``event engineering" first revealed the remarkable systematics of ``ridge"  like  rapidity separated azimuthal angle hadron correlations,  triggering debates regarding their initial state~\cite{Dumitru:2010iy,Dusling:2012iga}, or hydrodynamic origins~\cite{Bozek:2011if,Bozek:2012gr}.

Heavy flavor measurements add important elements to the discussion because the large quark masses provide a semi-hard scale to probe initial state dynamics. 
A compelling example of event engineered heavy flavor measurements in $p+p$ and $p+A$ collisions at RHIC and the LHC are ratios of their yields in high multiplicity events relative to minimum bias events. When plotted versus event activity,  the ratio of charged hadron multiplicity in rare relative to minimum bias events, many model dependencies cancel out.  In particular, because nonperturbative features of hadronization are likely the same for both rare and minimum bias events, ratios of heavy flavor multiplicities are sensitive primarily to short distance interactions of intermediate states.

The exciting possibility that event engineering may help distinguish between intermediate states can be quantified in the Nonrelativistic QCD (NRQCD)~\cite{Bodwin:1994jh} framework, wherein the inclusive differential cross-section of a heavy quarkonium state $Q$ in $p+p$ and $p+A$ collisions is expressed as
\begin{align}\label{eq:fact}
\frac{d\sigma_{Q}}{d^2\bm{p}_\perp}=\sum_\kappa \frac{d \sigma^\kappa_{Q\bar Q}}{d^2\bm{p}_\perp}\langle {\cal O}_\kappa^{Q}\rangle\,,
\end{align}
where $\kappa=\state{{2S+1}}{L}{J}{c}$ are quantum numbers of the
produced intermediate heavy quark pair, with $S$, $L$ and $J$ denoting its spin, orbital, and total angular momenta, respectively. The symbol $c$ denotes a color singlet (CS, $c=1$) or color octet (CO, $c=8$) state. The $d \sigma^\kappa$ are perturbative short distance coefficients for heavy quark pair production with  quantum numbers $\kappa$ and $\langle\mo^{Q}_\kappa\rangle$ are universal nonperturbative long distance matrix elements (LDMEs) . The LDMEs can for instance be extracted from data on quarkonium production at the Tevatron, and employed to make predictions for cross-sections at the RHIC and LHC.  While NRQCD is successful, an important puzzle is that the magnitude of the linear combination of the $\COaSz$ and $\COcPz$ LDMEs extracted from hadroproduction data~\cite{Ma:2010yw,Butenschoen:2010rq}  is larger than an upper bound set by BELLE $e^+ e^-$ data~\cite{Zhang:2009ym}. While this apparent breaking of universality may bring into question NRQCD factorization, we will show that event engineering offers a possible resolution to this puzzle.

In this work, we will show that the systematics of heavy flavor production in rare events in $p+p$  and $p+A$ collisions are sensitive to strongly correlated gluons in the colliding protons and nuclei. The dynamics of such configurations is controlled by an emergent semi-hard saturation scale $Q_s(x)$ in each of the colliding hadrons, where $x$ is the longitudinal momentum fraction carried by a parton in the hadron~\cite{Gribov:1984tu,Mueller:1985wy}.  Since $Q_s(x)$ grows with decreasing $x$, and increasing nuclear size, the interplay of the dynamics of hard and soft modes evolves with the changing energy and centrality of the collision.

A systematic framework to study gluon saturation is the Color Glass Condensate (CGC) effective field theory (EFT)~\cite{Gelis:2010nm,Kovchegov:2012mbw,Blaizot:2016qgz}. The cross-sections for the production of heavy quarkonia in the CGC EFT for hadron-hadron collisions were computed over a decade ago~\cite{Gelis:2003vh,Blaizot:2004wv,Tuchin:2004rb,Fujii:2005vj,Fujii:2006ab}. A more recent development is the CGC+NRQCD framework\footnote{See \cite{Dominguez:2011cy} for a specialized discussion.}~\cite{Kang:2013hta}, the novel element being that $d \sigma^\kappa$ in Eq.~(\ref{eq:fact}) is computed in the CGC EFT. There are several phenomenological studies of data from RHIC and LHC that employ these computations in $p+p$ and $p+A$ collisions~\cite{Fujii:2013gxa,Ma:2014mri,Ma:2015sia,Watanabe:2015yca,Ducloue:2015gfa,Fujii:2013yja,Fujii:2015lld,Ducloue:2016pqr,Fujii:2017rqa,Ma:2017rsu}.  High multiplicity configurations are approximated by increasing the  value of $Q_s(x)$ at the input large $x$ scale in both protons and nuclei in multiples of $Q_0^2=0.168\,\text{GeV}^2$, the initial saturation scale at $x=0.01$, determined from fits to the minimum bias $e+p$ DIS data~\cite{Tribedy:2010ab}. As also implemented in studies of ridge yields~\cite{Dusling:2013qoz,Dusling:2015rja,Schenke:2016lrs}, increasing the saturation scale in this manner captures the fluctuations of protons and nuclei into larger numbers of color charges in rare events. More systematic treatments of high multiplicity ``biased" color charge configurations are under development~\cite{Dumitru:2017cwt,Dumitru:2017ftq,Dumitru:2018iko}.

We will focus here\footnote{The $\Upsilon$ and open bottom computations require Sudakov resummation~\cite{Mueller:2012uf,Mueller:2013wwa,Qiu:2013qka,Qiu:2017xbx} and are beyond our scope here.} on measurements of  $D$ and $J/\psi$ mesons in high multiplicity $p+p$ and $p+A$ collisions~\cite{Adam:2015ota,Adam:2016mkz,Abelev:2012rz,Weber:2017hhm,Khatun:2017yic,Adamova:2017uhu,Ma:2015xta,Federicova:2017bmd}. The striking feature of the data is that the production yields of $D$ and $J/\psi$ in high multiplicity events are significantly enhanced relative to minimum bias events. Interestingly, in $p+p$ collisions, such growth is observed to be independent of collision energy. The models  proposed to explain their systematics include percolation models~\cite{Ferreiro:2012fb,Ferreiro:2015gea},  dipole models~\cite{Kopeliovich:2013yfa} and multiparton interaction models~\cite{Thakur:2017kpv}. All  these models approximate effects contained in the CGC EFT. Gluon saturation is included in the EPOS3 model~\cite{Werner:2013tya}, which also includes final state scattering effects.  As we will show, the CGC+NRQCD EFT can address detailed differential questions regarding heavy flavor production mechanisms and help resolve extant heavy flavor puzzles in collider experiments.

\begin{figure*}
\centering
\includegraphics[width=0.475\linewidth]{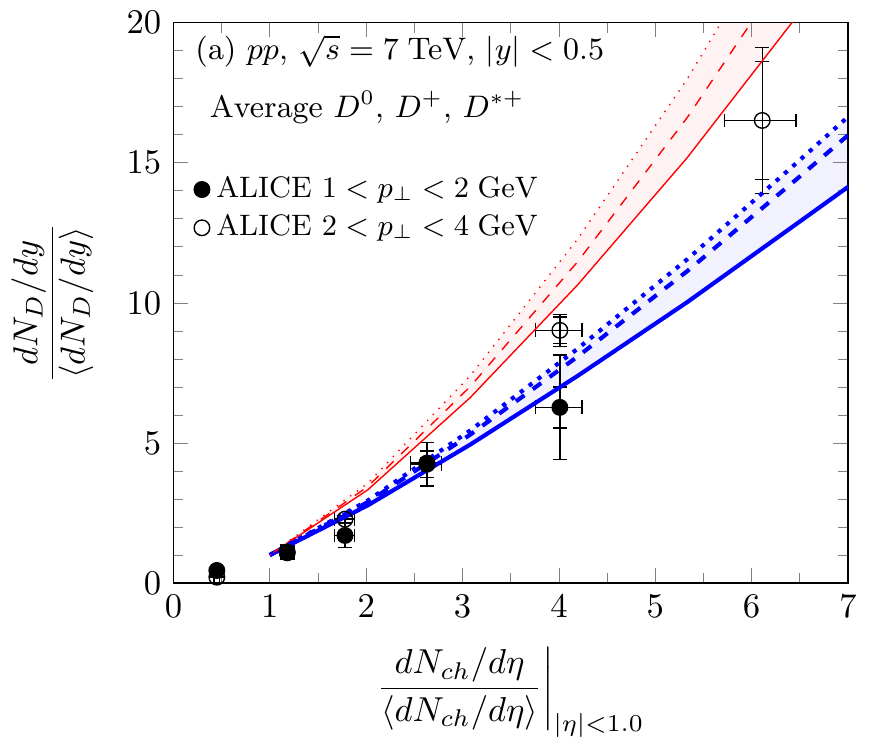}
\includegraphics[width=0.475\linewidth]{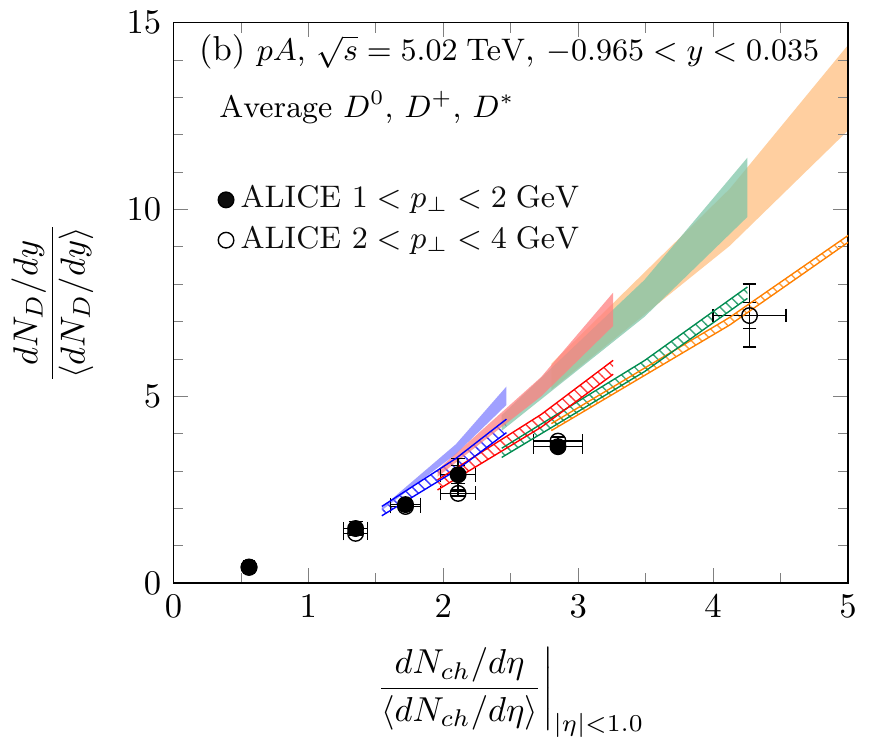}
\caption{(Color online)
(a): Relative yields of average $D$ ($D^0$, $D^+$, $D^{\ast+}$) as a function of relative multiplicity in $p+p$ collisions at the LHC. The thick (thin) curves are the results at $1<p_\perp<2\,\text{GeV}$ ($2<p_\perp<4\,\text{GeV}$)  using the BCFY (solid), BCFY+DGLAP (dashed), and KKKS (dotted) FFs, the bands representing the differences between these FF sets.  Data are from Ref.~\cite{Adam:2015ota}.
(b): Results in $p+A$ collisions. The hatched (filled) bands are the results at $1<p_\perp<2\,\textrm{GeV}$ ($2<p_\perp<4\,\textrm{GeV}$). The blue, red, green and orange bands all show model results for variations in the range $Q_{sp,0}^2 = \textrm{1--3} Q_0^2$ for $Q_{sA,0}^2=4,6,9,12 Q_0^2$ respectively with taking into account FF  uncertainties. Data are from Ref.~\cite{Adam:2016mkz}.}
\label{fig:D-Nch}
\end{figure*}

\section{Open flavor and Quarkonium production}{\label{section:open-flavor}}

We first consider the spin and color averaged inclusive cross-section $p+A(p)\rightarrow c (\bm{p}_c)+\bar c(\bm{q}_{\bar c})+X$, which  can be expressed in the CGC EFT as~\cite{Blaizot:2004wv}
\begin{align}
\frac{d \sigma_{c \bar{c}}}{d^2\bm{p}_{c\perp} d^2\bm{q}_{\bar c\perp} dy_c dy_{\bar c}}
=&
\frac{\alpha_s N_c^2 \pi R_{A}^2}{2(2\pi)^{10} d_A}
\int\limits_{\bm{k}_{2\perp},\bm{k}_\perp}
\frac{\varphi_{p,y_p}(\bm{k}_{1\perp})}
{k_{1\perp}^2}\non
&\times
\mathcal{N}_{Y}(\bm{k}_\perp)\mathcal{N}_{Y}(\bm{k}_{2\perp}-\bm{k}_\perp)\, \Xi\,,
\label{eq:xsection-kt-factorization-LN}
\end{align}
where $\int_{\bm{k}_\perp}=\int d^2\bm{k}_\perp$, $k_{1\perp}=|\bm{k}_{1\perp}|$, $d_A=N_c^2-1$, with
$\bm{p}_{c\perp}$ ($\bm{q}_{\bar c\perp}$) and $y_c$ ($y_{\bar c}$), the transverse momentum
and rapidity respectively of the produced charm (anti-charm) quarks. Further, $y_p=\ln(1/x_1)$ and $Y=\ln(1/x_2)$, where
$x_{1,2}=(\sqrt{m_c^2+p_{c\perp}^2}e^{\pm y_c}+\sqrt{m_c^2+q_{\bar c\perp}^2}e^{\pm y_{\bar c}})/\sqrt{s}$, denote the
longitudinal momentum fractions of the interacting gluons in the projectile and target respectively.
The expression for the hard scattering matrix element $\Xi$ is listed in Appendix~\ref{section:appendixA}.  The unintegrated gluon distribution function (UGDF) of the projectile proton
$\varphi_{p,y_p}(\bm{k}_\perp)$ is defined as~\cite{Ma:2014mri}
\begin{align}
\varphi_{p,y_p}(\bm{k}_{\perp}) = \pi R_p^2\, \frac{N_ck_{\perp}^2}{4\alpha_s} \mathcal{N}^A_{y_p}(\bm{k}_{\perp})\,.
\label{eq:proton-wavefn}
\end{align}
Here $\pi R_p^2$ ($\pi R_A^2$) is the transverse area occupied by gluons in the proton (nucleus) and $\mathcal{N}^A_{y_p}(\bm{k}_{\perp})=\int d^2l_\perp/(2\pi)^2\mathcal{N}_{y_p}(\bm{k}_{\perp}-\bm{l}_{\perp}) \mathcal{N}_{y_p}(\bm{l}_{\perp})$. The fundamental dipole amplitude is given by
\begin{align}
\mathcal{N}_{y_p(Y)}(\bm{k}_{\perp})=&\int d^2\bm{r}_\perp e^{-i\bm{k}_{\perp}\cdot \bm{r}_\perp}\non
&\times
\frac{1}{N_c}\left<\mathrm{Tr}\left[V_F(\bm{r}_\perp)V_F^\dagger(\bm{0}_\perp)\right]\right>_{y_p(Y)}\,,
\end{align}
where $V_F(\bm{r}_\perp)$ ($V_F^\dagger(\bm{0}_\perp)$) is the  fundamental Wilson line in the amplitude (complex conjugate amplitude)  representing multiple scattering of the quark with background fields at the position $\bm{r}_\perp$ ($\bm{0}_\perp$). Note that $\left<\cdots\right>_{y}$ here corresponds to leading log x resummation in the CGC EFT and must not be confused with the LDMEs expectation value in Eq.~(\ref{eq:fact}).

The differential cross section for $D$ meson production is then given by
\begin{align}
\frac{d \sigma_{D}}{d^2\bm{p}_{D\perp} dy}
=
&\int_{z_{min}}^1 dz\frac{D_{c\rightarrow D}(z)}{z^2}\non
&\times
\int dy_{\bar c} \int_{\bm{q}_{\bar c\perp}}\frac{d \sigma_{c \bar{c}}}{d^2\bm{p}_{c\perp} d^2\bm{q}_{\bar c\perp}dy dy_{\bar c}}\,,
\label{eq:fragmentation}
\end{align}
where $D_{c\rightarrow D}(z)$ is the fragmentation function (FF) for $D^0$, $D^+$, $D^{\ast+}$ mesons, with $z =p_{D\perp}/p_{c\perp}$.
It satisfies $\int dz D_{c\rightarrow D}(z)=Br(c\rightarrow D)$; the branching ratio $Br(c\rightarrow D)$  for the transition from $c$ to $D$, in turn, satisfies $\sum_X Br(c\rightarrow X)=1$ with $X$ denoting all heavy flavor hadrons. We will employ here the BCFY~\cite{Braaten:1994bz} and KKKS~\cite{Kneesch:2007ey} FFs; key details are discussed in Appendix~\ref{section:appendixB}.

The color singlet ($\kappa=\CScSa$) channel contribution of $J/\psi$ production cross-section in the CGC+NRQCD framework can be expressed as~\cite{Ma:2014mri}
\begin{align}
\frac{d \sigma_{c \bar{c},\rm{CS}}^{\kappa}}{d^2\bm{p}_{\perp} dy}
&=
\frac{\alpha_s\pi R_A^2}{(2\pi)^{9} d_A}
\int\limits_{\bm{k}_{2\perp},\bm{k}_\perp,\bm{k}_\perp^\prime}
\frac{\varphi_{p,y_p}(\bm{k}_{1\perp})}
{k_{1\perp}^2}
\non
\times&\,
{\cal N}_Y(\bm{k}_\perp){\cal N}_Y(\bm{k}_\perp^\prime){\cal N}_Y(\bm{k}_{2\perp}-\bm{k}_\perp-\bm{k}_\perp^\prime)
\,
{\cal G}_{1}^\kappa\,,
\label{eq:xsection-kt-factorization-LN-CS}
\end{align}
and the color octet (CO) intermediate states are written as
\begin{align}
\frac{d \sigma_{c \bar{c},\rm{CO}}^{\kappa}}{d^2\bm{p}_{\perp} dy}
=&
\frac{\alpha_s\pi R_A^2}{(2\pi)^{7} d_A}
\int\limits_{\bm{k}_{2\perp},\bm{k}_\perp}
\frac{\varphi_{p,y_p}(\bm{k}_{1\perp})}
{k_{1\perp}^2}
\non
&\times
{\cal N}_Y(\bm{k}_\perp){\cal N}_Y(\bm{k}_{2\perp}-\bm{k}_\perp)
\,
{\Gamma}_{8}^\kappa\,.
\label{eq:xsection-kt-factorization-LN-CO}
\end{align}
The  hard matrix elements ${\cal G}_{1}^\kappa$ and ${\Gamma}_{8}^\kappa$ are given in Appendix~\ref{section:appendixA}. Note that $x_{1,2}=\sqrt{(2m_c)^2+p_\perp^2}e^{\pm y}/\sqrt{s}$ in $y_p$ and $Y$ where $m_c = m_{J/\psi}/2$.  Since Eq.~\eqref{eq:xsection-kt-factorization-LN-CS} has a cubic dependence on  ${\cal N}_Y$,  while Eq.~\eqref{eq:xsection-kt-factorization-LN-CO} has only a quadratic dependence, it is evident that the short distance CS and CO cross-sections have different dependencies on the dynamics of saturated gluons in protons and nuclei.

We will compare the NRQCD results employing the above expressions with the $J/\psi$ cross-section computed in the Improved Color Evaporation Model (ICEM)~\cite{Ma:2016exq}. The differential cross section for $J/\psi$ production in the CGC+ICEM framework is given by
\begin{align}
\frac{d\sigma_{J/\psi}}{d^2\bm{p}_\perp dy}=&F_{J/\psi}\int_{m_{J/\psi}}^{2m_D}dM \left(\frac{M}{m_{J/\psi}}\right)^2 \int\limits_{0}^{\sqrt{\frac{M^2}{4}-m_c^2}} \hspace{-10pt} d\tilde{q}
\int\limits_0^{2\pi}d\phi \non
&\times J\,\frac{d\sigma_{c\bar{c}}}{d^2\bm{p}_{c\perp} d^2\bm{q}_{\bar c\perp} dy_c dy_{\bar{c}}}\,,
\label{eq:ICEM}
\end{align}
where $J=\tilde{q} \sqrt{M^2+{p}^2_\perp}/\left[M\omega_c\omega_{\bar c}|\sinh(y_c-y_{\bar{c}})|\right]$ with $\omega_{c}=\sqrt{m_c^2+{p}_{c\perp}^2}$ and $\omega_{\bar c}=\sqrt{m_c^2+{q}_{\bar c\perp}^2}$. Here $M$ is the invariant mass of the $c\bar c$. $\tilde{q}$ and $\phi$ are respectively the relative momentum and angle between $c$ and $\bar c$ in the $c\bar c$ pair rest frame~\cite{Fujii:2006ab}. $F_{J/\psi}$ represents the nonperturbative transition probability from the $c\bar c$ pair to the $J/\psi$ meson.  The principal difference between the ICEM and the conventional CEM~\cite{Fritzsch:1977ay,Gluck:1977zm,Barger:1979js} is that the $J/\psi$'s transverse momentum differs from the pair's transverse momentum $\bm{p}_\perp^\prime$: $\bm{p}_\perp=(m_{J/\psi}/M) \bm{p}_\perp^\prime$. In our computations, we will use $m_{J/\psi}=3.1\,{\rm GeV}$ and $2\,m_D=3.728\,{\rm GeV}$.

\section{Results for $D$-meson and $J/\psi$ production}\label{section:results}

With the expressions in Eqs.~\eqref{eq:fragmentation},~\eqref{eq:xsection-kt-factorization-LN-CS},~\eqref{eq:xsection-kt-factorization-LN-CO} and \eqref{eq:ICEM}, we can simultaneously study $D$-meson and $J/\psi$ production with increasing event activity, as represented by the  inclusive charged hadron multiplicity. The latter is computed in a $k_\perp$ factorized approximation to the CGC EFT~\cite{Tribedy:2010ab,Tribedy:2011aa,Albacete:2012xq} as shown in Appendix~\ref{section:appendixC}. The dynamical ingredients in all the computations are the UGDs in the projectile and the target. Therefore fixing these, and their energy evolution (see Appendix~\ref{section:appendixD}) from single inclusive production provides significant predictive power.  In Appendix~\ref{section:appendixE}, we present numerical results for the charged hadron multiplicity. As shown there, these initial scales $Q^2_{sp,0}$ ($Q^2_{sA,0}$) at  $x=0.01$ for protons (nuclei) that enter into the UGDs are well constrained by the data on $\langle p_\perp\rangle$ versus $dN_{\rm ch}/d\eta$ of charged hadrons. For the event engineering studies, the UGDs are obtained by varying  $Q^2_{sp,0}$ ($Q^2_{sA,0}$) within a range of 1--3 (4--12) times their corresponding minimum bias values $(Q_0^2=0.168\,\text{GeV}^2)$.

With $Q_{sp,0}$ and $Q_{sA,0}$ thereby constrained, the UGDs can be used to compute the isospin averaged $D$-meson cross-section. Figure~\ref{fig:D-Nch} compares our model prediction to the mid-rapidity LHC high multiplicity data in both $p+p$ and $p+A$ collisions, normalized to the  minimum bias value, versus $dN_{\rm ch}/d\eta$ likewise normalized to its minimum bias value. As is clear from Eqs.~\eqref{eq:xsection-kt-factorization-LN}--\eqref{eq:fragmentation}, the ratio plotted on the y-axis is fairly insensitive to uncertainties arising from choice of fragmentation functions, proton and nuclear size, and the coupling constant $\alpha_s$. Likewise, the ratio on the $x$-axis minimizes nonperturbative uncertainties from geometry effects in both protons and nuclei. The agreement with $p+p$ data at $\sqrt{s}=7$ TeV shown in Fig.~\ref{fig:D-Nch} (a) is remarkably good for both $p_\perp$ windows. The experimental error bars are however large for the rarest events. Figure~\ref{fig:D-Nch} (b) shows the model comparison to LHC $p+A$ data at $\sqrt{s}=5.02$ TeV/nucleon. While model agreement with data in the $1<p_\perp < 2$ GeV window is quite good, it overshoots data for $2<p_\perp<4$ GeV though it has the same qualitative trend. Because one varies both $Q_{sp,0}$ and $Q_{sA,0}$, there is room for finetuning. Appendix~\ref{section:appendixF} shows that $D$-meson $p_\perp$ distributions for minimum bias events are well reproduced out to $p_\perp\sim 5$ GeV in both $p+p$ and $p+A$ collisions.

\begin{figure}
\centering
\includegraphics[width=0.95\linewidth]{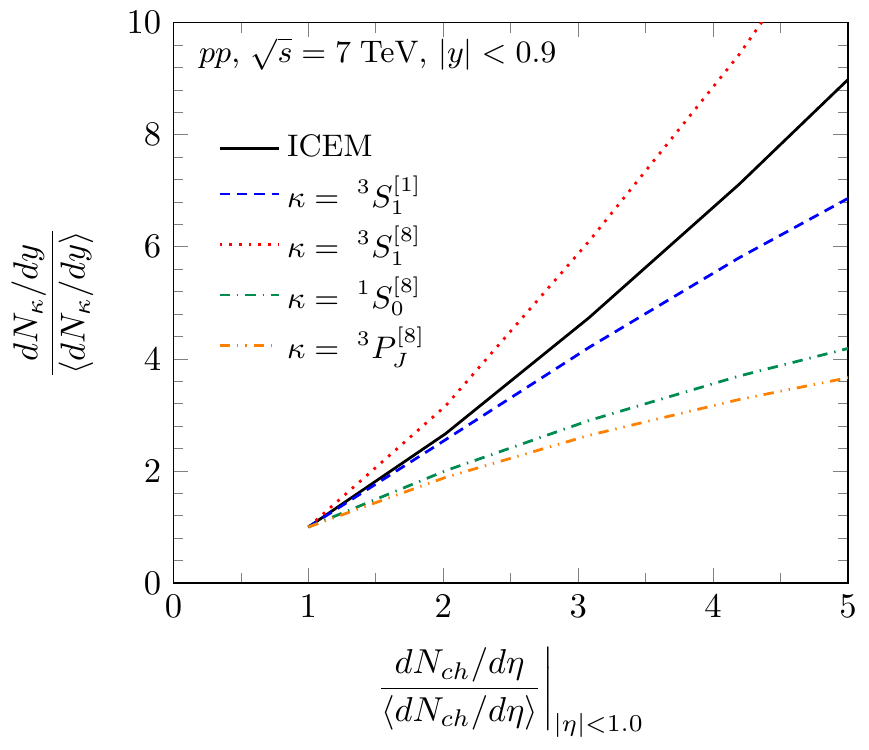}
\caption{(Color online)
Relative yield of $J/\psi$ production as a function of relative multiplicity in $p+p$ collisions at mid rapidity at the LHC. The solid line is obtained in the CGC+ICEM model. Other lines correspond to contributions from different intermediate states in the CGC+NRQCD framework. 
}
\label{fig:Jpsi-pp-multi}
\end{figure}

The very same UGDs are used to compute $J/\psi$ production. Remarkably, the relative contribution of $d\sigma^\kappa$ for each $\kappa$ changes with increasing event activity. Figure~\ref{fig:Jpsi-pp-multi} shows that relative yield of $^3S_1^{[8]}$ is larger than the other channels for all $dN_{\rm ch}/d\eta$, and it increases significantly with increasingly rare events. This implies that a very rapid growth in $J/\psi$ production in rare events in the $^3S_1^{[8]}$  channel relative to minimum bias. The growth in the contributions of the $^1S_0^{[8]}$  and $^3P_J^{[8]}$ channels are relatively much smaller. This enhanced contribution of the short distance contributions in the $^3S_1^{[8]}$ channel suggests the LDMEs of the $^1S_0^{[8]}$ and  $^3P_J^{[8]}$ channels could potentially be smaller. This may provide a way forward in reconciling the LDMEs extracted from hadroproduction with the 
universality requirement extracted from BELLE $e^+ e^-$ data, hence providing a possible resolution of the NRQCD puzzle mentioned previously.

\begin{figure*}
\centering
\includegraphics[width=0.475\linewidth]{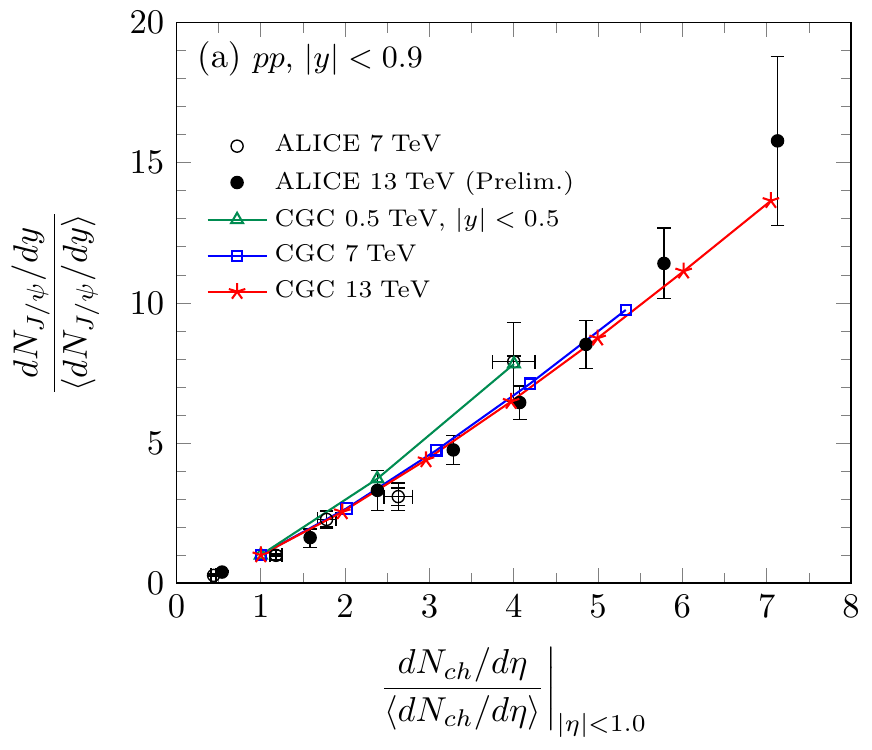}
\includegraphics[width=0.475\linewidth]{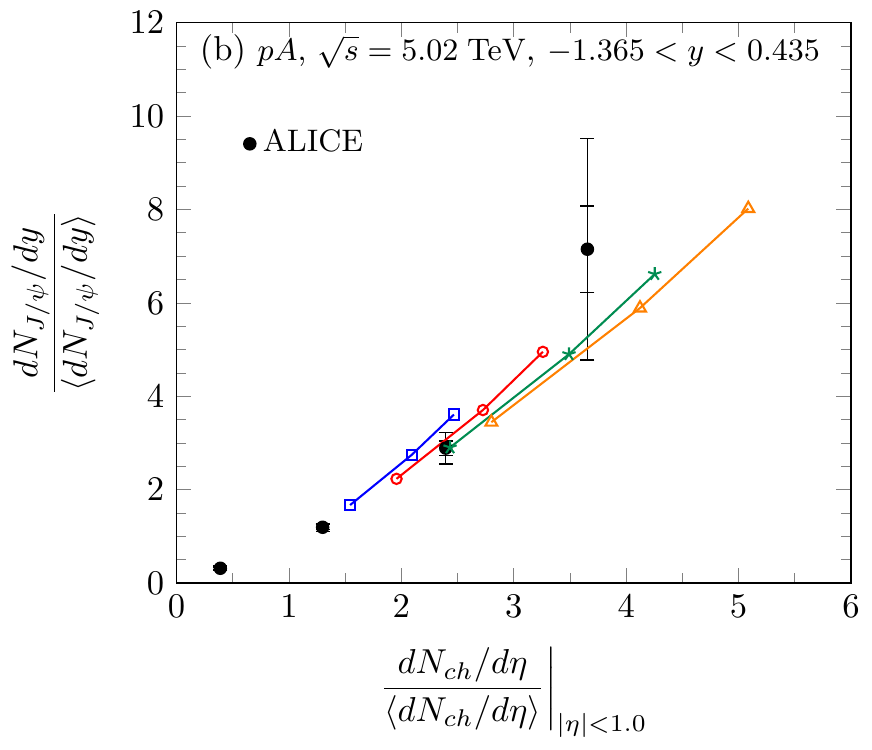}
\caption{(Color online)
(a): $N_{ch}$ dependence of $J/\psi$ production in $p+p$ collisions at mid rapidity at $\sqrt{s}=7\,\text{TeV}$, 13\,TeV, and 0.5\,TeV in the CGC+ICEM model. Data at $\sqrt{s}=7$\,TeV from Ref.~\cite{Abelev:2012rz}. Preliminary $\sqrt{s}=13$\,TeV data are from Refs.~\cite{Weber:2017hhm,Khatun:2017yic}.
(b): Results for $J/\psi$ production vs $N_{ch}$ in $p+A$ collisions at $\sqrt{s}=5.02\,$TeV  in the CGC+ICEM model. Data are from Ref.~\cite{Adamova:2017uhu}.}
\label{fig:Jpsi-Nch-pp+pA}
\end{figure*}

The relative large $^3S_1^{[8]}$ contribution suggests that the simpler ICEM model where, gluon fragmentation through this channel dominates, may be sufficient to describe $J/\psi$ production and we will do so in the following. In future, we will study rare events directly in the CGC+NRQCD framework. Figure~\ref{fig:Jpsi-Nch-pp+pA} (a) shows that the data on ratios of the $J/\psi$ cross-section in $p+p$ collisions is $\sqrt{s}$--independent. In the CGC, as seen previously for ridge yields~\cite{Dusling:2015rja}, the energy dependence of cross-sections is controlled by $Q_s(x)$, which also governs the charged hadron multiplicity; events at different energies with the same $Q_s$ are therefore identical. Figure~\ref{fig:Jpsi-Nch-pp+pA} (a)  predicts that
RHIC $p+p$ data at $\sqrt{s}=0.5$ TeV will conform to this expectation. In Fig.~\ref{fig:Jpsi-Nch-pp+pA} (b), we compare the CGC+ICEM model to data in $p+A$ collisions. Since many nonperturbative uncertainties cancel in these ratios, the agreement with both $p+p$ and $p+A$ data demonstrates that the CGC EFT captures key features of the short distance cross-sections.

\section{Summary}\label{section:summary}

We outlined the potential of event engineered heavy flavor measurements to uncover the dynamics of rare parton configurations at collider energies. 
Our CGC EFT studies  suggest that the short distance dynamics in such events requires saturation scales that are an order of magnitude greater than those in minimum bias events. On the one hand, these harder scales suggest that the weak coupling CGC framework is more reliable for rare events. On the other hand, the treatment of rare multiplicity biased configurations is significantly more complex than computations developed to study minimum bias configurations and demands further theoretical development.

Our work further illustrates the potential of event engineering to distinguish between intermediate states with differing quantum numbers that contribute to the hadronization of quarkonia. The finding that the hadronization contribution of the $^3S_1^{[8]}$ state to $J/\psi$ production grows rapidly suggests the growing importance of hard gluon fragmentation in $J/\psi$ hadronization. As noted, this result may provide an important clue in resolving the universality requirements on LDMEs from BELLE $e^+e^-$ data, thereby possibly resolving a puzzle between the magnitudes of the LDMEs extracted from hadron collision data relative to $e^+e^-$ data.

A systematic theoretical uncertainty is that the dilute-dense approximation to CGC EFT we employ is valid only when $Q_{s,{\rm proj.}}/k_{\perp, {\rm proj.}} < Q_{s,{\rm target}}/k_{\perp,{\rm target}}$. The full ``dense-dense" EFT computation is beyond the scope of present computations; these are beginning to be quantified~\cite{Tanji:2017xiw}. This systematic uncertainty is reduced at forward rapidities in $p+p$ collisions and at both central and forward rapidities in $p+A$ collisions. The ratios considered mitigate these uncertainties; further, the requirement that we reproduce charged particle multiplicities is a powerful constraint. Our results for the $J/\psi$ ratios at forward rapidities are presented in Appendix~\ref{section:appendixF}. Within the uncertainties noted, we find good agreement with data. The model, with the  parameters thus fixed, can for example be compared to data on $J/\psi$-hadron correlations at the LHC~\cite{Acharya:2017tfn}.

Finally, a source of systematic uncertainty in our computation we have not discussed is the  possible role of higher twist fragmentation contributions at low $p_\perp$. The short distance hard matrix elements ensure any such contribution is suppressed by $\alpha_s(m_Q)$. Such higher order contributions, as well as other $\alpha_s$ suppressed contributions to the matrix elements are not included in our treatment. Our framework however can be systematically improved in future to include such effects.

%%%%%%%%%%%%%%%%%%%%%%%%%%%%%%%%%%%%%%%%%%%%%%%%
\subsection*{Acknowledgement}

K.W. is supported by Jefferson Science Associates, LLC under U.S. DOE Contract No.\ DE-AC05-06OR23177 and U.S. DOE Grant No.\ DE-FG02-97ER41028. P.T. and R.V. are supported by the U.\ S.\ Department of Energy Office of Science, Office of Nuclear Physics, under contracts No.\ DE-SC0012704. This work is part of and supported by the DFG Collaborative Research Centre ``SFB 1225 (ISOQUANT)''. We would like to thank Anton Andronic, Juergen Berges, Peter Braun-Munzinger, Rongrong Ma, Silvia Masciocchi, Jianwei Qiu, Klaus Reygers, Lijuan Ruan, and Johanna Stachel for useful comments. R.V. would in particular like to thank Adrian Dumitru and Vladimir Skokov for sharing their deep insights on a theoretical framework for multiplicity biased events. Not least, we would like to thank Tomasz Stebel for a careful reading of the manuscript and for helpful suggestions on improving the text.

\appendix

%%%%%%%%%%%%%%%%%%
%Inclusive hadron

\section{Hard Matrix Elements}
\label{section:appendixA}

\subsection{Hard Matrix element in $c\bar{c}$ production}
The explicit expression for
$\Xi$ in Eq.~\eqref{eq:xsection-kt-factorization-LN} for $D$-meson production and in Eq.~\eqref{eq:ICEM} for $J/\psi$ production is given by $\Xi=\Xi^{q\bar{q},q\bar{q}}+\Xi^{q\bar{q},g}+\Xi^{g,g}$, where
\begin{align}
&\Xi^{q\bar{q},q\bar{q}}
= \frac{32 p_c^+q_{\bar c}^+(m^2+a_\perp^2)(m^2+b_\perp^2)}{[2p_c^+(m^2+a_\perp^2)+2q_{\bar c}^+(m^2+b_\perp^2)]^2},\\
&\Xi^{q\bar{q},g}
= \frac{16}{2(m^2+p_c\cdot q_{\bar c})[2p_c^+(m^2+a_\perp^2)+2q_{\bar c}^+(m^2+b_\perp^2)]}\non
&\times\!
\bigg[(m^2+\bm{a}_\perp\!\!\cdot\!\bm{b}_\perp)\{q_{\bar c}^+C\!\cdot\!p_c+p_c^+C\!\cdot\!q_{\bar c}-C^+(m^2+p_c\!\cdot\!q_{\bar c})\}\non
&+C^+\{(m^2+\bm{b}_\perp\!\!\cdot\!\bm{q}_{\bar c\perp})(m^2-\bm{a}_\perp\!\!\cdot\!\bm{p}_{c\perp})\non
&\hspace{50pt}-(m^2+\bm{a}_\perp\!\!\cdot\!\bm{q}_{\bar c\perp})(m^2-\bm{b}_\perp\!\!\cdot\!\bm{p}_{c\perp})\}\non
&+p_c^+\{\bm{a}_\perp\!\!\cdot\!\bm{C}_\perp(m^2+\bm{b}_\perp\!\!\cdot\!\bm{q}_{\bar c\perp})-\bm{b}_\perp\!\!\cdot\!\bm{C}_\perp(m^2+\bm{a}_\perp\!\!\cdot\!\bm{q}_{\bar c\perp})\}\non
&+q_{\bar c}^+\{\bm{a}_\perp\!\!\cdot\!\bm{C}_\perp(m^2-\bm{b}_\perp\!\!\cdot\!\bm{p}_{c\perp})-\bm{b}_\perp\!\!\cdot\!\bm{C}_\perp(m^2-\bm{a}_\perp\!\!\cdot\!\bm{p}_{c\perp})\}\bigg],\\
&\Xi^{g,g}
= \frac{4\left[2(p_c\cdot C)(q_{\bar c}\cdot C)-(m^2+p_c\cdot q_{\bar c})C^2\right]}{4(m^2+p_c\cdot q_{\bar c})^2}\,.
\end{align}
In the above, $\bm{a}_\perp=\bm{q}_{{\bar c}\perp}-\bm{k}_\perp$ and $\bm{b}_\perp=\bm{q}_{{\bar c}\perp}-\bm{k}_\perp-\bm{k}_{1\perp}$. The Lipatov vertex $C^\mu$ that appears here, can be written in component form as $C^+=\;p_c^++q_{\bar c}^+-\frac{k_{1\perp}^2}{p_c^-+q_{\bar c}^-}$, $C^-=\;\frac{k_{2\perp}^2}{p_c^++q_{\bar c}^+}-(p_c^-+q_{\bar c}^-)$, and  $\bm{C}_\perp=\;\bm{k}_{2\perp}-\bm{k}_{1\perp}$.

\subsection{NRQCD}
For the color singlet $^3\!S_1$ channel, ${\cal G}_{1}$ reads~\cite{Ma:2014mri}
\begin{align}
{\cal G}_{1}^{^3\!S_1}=\frac{k_{1\perp}^2(k_{1\perp}^2+4m^2)}{12m}\left(\frac{1}{X_{l_\perp}}-\frac{1}{X_{l^\prime_\perp}}\right)^2
\end{align}
where $X_{l_\perp}\equiv l_\perp^2+{k_{1\perp}^2}/{4}+m^2$,
$X_{l_\perp^\prime}\equiv l_\perp^{\prime2}+{k_{1\perp}^2}/{4}+m^2$,
with
$\bm{l}_\perp=\bm{k}_\perp-{\bm{k}_{2\perp}}/{2}$ and
$\bm{l}_\perp^\prime=\bm{k}_\perp^\prime-{\bm{k}_{2\perp}}/{2}$.
Note here $\bm{k}_{2\perp}=\bm{p}_\perp-\bm{k}_{1\perp}$ due to momentum conservation at LO.

For the color octet channels, ${\Gamma}_{8}^\kappa$ reads~\cite{Kang:2013hta}
\begin{align}
&{\Gamma}_{8}^{^1\!S_0^{[8]}}= \frac{2[k_{1\perp}^2l_\perp^2-(\bm{k}_\perp\cdot \bm{l}_\perp)^2]}{mX_l^2},\\
&{\Gamma}_{8}^{^3\!S_1^{[8]}}= \frac{2k_{1\perp}^2(k_{2\perp}^2+4m^2)}{3m^2(p_\perp^2
	+4m^2)}\non
&-\frac{4k_{1\perp}^2(k_{2\perp}^2+\bm{k}_{1\perp}\cdot \bm{p}_\perp+4m^2)}{3mX_{l_\perp}(p_\perp^2+4m^2)}
+\frac{k_{1\perp}^2(k_{1\perp}^2+4m^2)}{6mX_{l_\perp}^2},\\
&{\Gamma}_{8}^{^3\!P_J^{[8]}}= \frac{4k_{1\perp}^2l_\perp^2-2(\bm{k}_{1\perp}\cdot\bm{l}_\perp)^2}{9m^3X_{l_\perp}^2}\non
+&\frac{2k_{1\perp}^2(\bm{k}_{1\perp}\cdot\bm{l}_\perp)(\bm{k}_{2\perp}\cdot\bm{l}_\perp)-8m^2\left[k_{1\perp}^2l_\perp^2-(\bm{k}_{1\perp}\cdot\bm{l}_\perp)^2\right]}{9m^3X_{l_\perp}^3}\non
+&\frac{k_{1\perp}^2(k_{1\perp}^2+4m^2)\left[(\bm{k}_{2\perp}\cdot \bm{l}_\perp)^2+4m^2l_\perp^2\right]}{18m^3X_{l_\perp}^4}.
\end{align}

\section{$D$-meson fragmentation functions}
\label{section:appendixB}

\begin{figure*}
	\centering
	\includegraphics[width=1\linewidth]{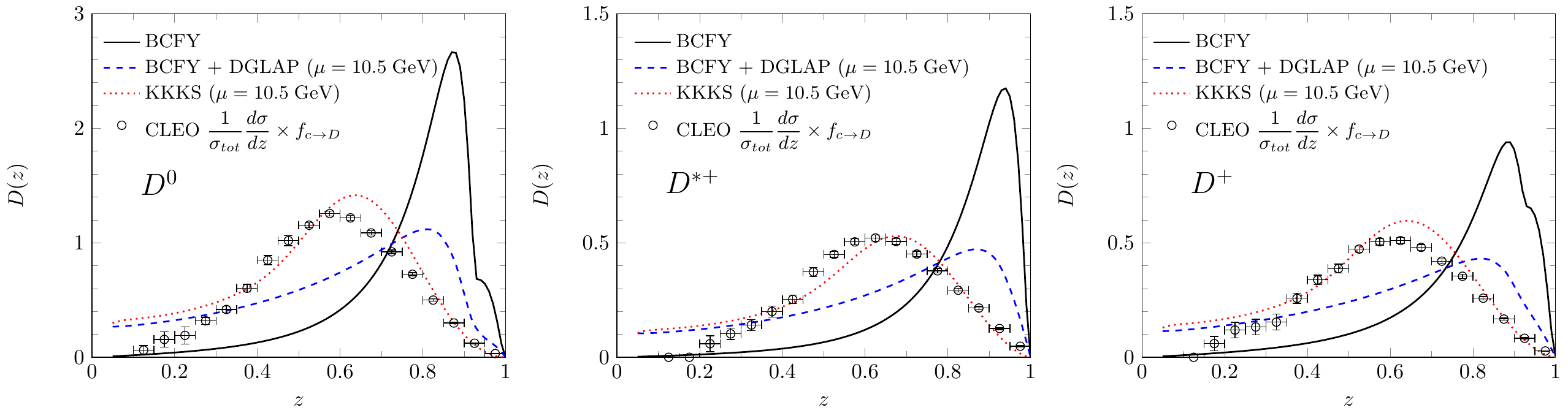}
	\caption{(Color online)
		Comparisons between the BCFY and KKKS FFs for $D^0$, $D^{\ast+}$, and $D^+$ mesons production. Solid curves are obtained directly from Eqs.~\eqref{eq:BCFY-D0}--\eqref{eq:BCFY-Dstar} with $r=0.1$. Blue dashed curves are obtained by putting those BCFY FFs in the DGLAP equation. $\mu$ is evolved from $\mu=1.5$\;GeV to $10.5$\;GeV. The KKKS FFs are shown as red dotted curves. CLEO $e^+e^-$ data at $\sqrt{s}=10.5$\;GeV are from~\cite{Artuso:2004pj}. $\sigma_{tot}^{D^0}=1550$\;pb, $\sigma_{tot}^{D^{\ast+}}=575$\;pb, and $\sigma_{tot}^{D^+}=640$\;pb are taken from~\cite{Artuso:2004pj}. Branching fractions are chosen as $f(c\rightarrow D^0)=0.560$, $f(c\rightarrow D^{\ast+})=0.233$, and $f(c\rightarrow D^+)=0.238$~\cite{Barate:1999bg}.
	}\label{fig:dff-data-comparison}
\end{figure*}

We will discuss here heavy-quark fragmentation functions (FF) that provide different $z$-distributions for pseudoscalar mesons and vector mesons. 
We consider specifically the Braaten-Cheung-Fleming-Yuan (BCFY) FF~\cite{Braaten:1994bz} and the  Kneesch-Kniehl-Kramer-Schienbein (KKKS) FF~\cite{Kneesch:2007ey}. Consider the BCFY FF first, following Refs.~\cite{Cacciari:2012ny,Cacciari:2003zu}, we will set the different FF for $D^0$, $D^+$, and $D^\ast$ production to be
\begin{align}
D_{c\rightarrow D^0}(z;r)&=0.168 D^{(P)}_\text{BCFY}(z;r)+0.39\tilde{D}^{(V)}_\text{BCFY}(z;r),\label{eq:BCFY-D0}\\
D_{c\rightarrow D^+}(z;r)&=0.162 D^{(P)}_\text{BCFY}(z;r)+0.07153\tilde{D}^{(V)}_\text{BCFY}(z;r),\label{eq:BCFY-Dp}\\
D_{c\rightarrow D^\ast}(z;r)&=0.233 D^{(V)}_\text{BCFY}(z;r),\label{eq:BCFY-Dstar}
\end{align}
where the original BCFY FFs are given by~\cite{Braaten:1994bz}
\begin{align}
D&^{(P)}_\text{BCFY}(z;r)=N\frac{rz(1-z)^2}{[1-(1-r)z]^6}\bigg[6-18(1-2r)z\non
&\;+(21-74r+68r^2)z^2-2(1-r)(6-19r+18r^2)z^3\non
&\;+3(1-r)^2(1-2r+2r^2)z^4\bigg],\\
D&^{(V)}_\text{BCFY}(z;r)=3N\frac{rz(1-z)^2}{[1-(1-r)z]^6}\bigg[2-2(3-2r)z\non
&\;+3(3-2r+4r^2)z^2-2(1-r)(4-r+2r^2)z^3\non
&\;+(1-r)^2(3-2r+2r^2)z^4\bigg].
\end{align}
$N$ is determined analytically from $\int_0^1 dz D^{(P,V)}_\text{BCFY}(z;r)=1$.
Here $\tilde{D}_\text{BCFY}^{(V)}$ describes $D^\ast$ production involving the effect of the $D^\ast$ decay into $D$, and reads
\begin{align}
\tilde{D}^{(V)}_\text{BCFY}(z;r)=\theta\left(\frac{m_D}{m_{D^\ast}}-z\right)D^{(V)}_\text{BCFY}\left(\frac{m_{D^\ast}}{m_D}z;r\right)\frac{m_{D^\ast}}{m_D}.
\end{align}
We shall fix $m_D=(m_{D^0}+m_{D^\pm})/2=1.867$\;GeV and $m_{D^\ast}=(m_{D^{\ast0}}+m_{D^{\ast\pm}})/2=2.009$\;GeV.
$r$ is a single nonperturbative parameter and can be interpreted as the ratio of the constituent mass of the light quark to the mass of the heavy meson like $r\sim(m_D-m_c)/m_D$. One can easily estimate $r={\mathcal O}(0.1)$. $z$ distribution of Eqs.~\eqref{eq:BCFY-D0}--\eqref{eq:BCFY-Dstar} are shown as solid curves in Fig.~\ref{fig:dff-data-comparison}.

The renormalization scale ($\mu$) dependence of the BCFY FFs can be implemented by solving the DGLAP evolution equation. Figure~\ref{fig:dff-data-comparison} also displays the DGLAP evolution of the BCFY FFs by setting \eqref{eq:BCFY-D0}--\eqref{eq:BCFY-Dstar} as initial conditions and evolving $\mu$ from 1.5\;GeV to 10.5\;GeV. Clearly, the DGLAP evolution significantly modifies the initial BCFY FFs.

Turning now to the KKKS FF, in the KKKS set\footnote{Numerical points of the KKKS FF as well as other FF set are available online thanks to~\cite{ffgenerator}.},  the $\mu$ dependence of the FFs for $D$-mesons was again taken into account through DGLAP evolution. As to initial conditions, the functional form $D_{c\rightarrow D}(z,\mu_0)=Nz^{-(1+\gamma^2)}(1-z)^a e^{-\gamma^2/z}$ is set at $\mu_0=1.5$\;GeV. All the input parameters $N$, $a$, $\gamma$ are determined by global fitting of all available $e^+e^-$ data. In Fig.~\ref{fig:dff-data-comparison}, the KKKS FFs at $\mu=10.5$\;GeV are compared to the BCFY FFs together with CLEO $e^+e^-$ data~\cite{Artuso:2004pj}. The data comparisons obviously prefer the KKKS FFs to describe $e^+e^-$ data, although one must keep in mind that the data are normalized cross-sections for $D$-mesons production, not heavy quark FFs themselves. Indeed, both the BCFY FF with the DGLAP evolution and the KKKS FF overshoot the data points at lower $z$ because we do not convolute hard scattering part with the FFs here for simplicity. If we take into account hard scattering part correctly, the KKKS FFs should agree with the data~\cite{Kneesch:2007ey}.

\section{Inclusive hadron production}
\label{section:appendixC}

We will review here charged hadron production in $p+p$ and $p+A$ collisions in the CGC framework~\cite{Tribedy:2010ab,Tribedy:2011aa,Albacete:2012xq}. The  differential cross-section for inclusive gluon production in $p+A$ collisions ($p+A\rightarrow g(\bm{p}_g)+X$) in the $k_\perp$-factorization formula at LO~\cite{Kovchegov:2001sc,Blaizot:2004wu} is given by
\begin{align}
\frac{d\sigma_{p+A\rightarrow g+X}}{d^2\bm{p}_{g\perp} dy}=&\frac{\alpha_s \hat{K}_{b}}{(2\pi)^3\pi^3 C_F}\frac{1}{{p}_{g\perp}^2}\int d^2\bm{k}_\perp \varphi_{p,y_p}(\bm{k}_\perp)\non
&\times\varphi_{A,Y}(\bm{p}_{g\perp}-\bm{k}_\perp)
\label{eq:gluon-kt}
\end{align}
where $k_\perp\leq p_{g\perp}$. Now in $y_{p}$ and $Y$, one should read $x_{1,2}=p_{g\perp}e^{\pm y}/\sqrt{s}$. The impact parameter dependence is encoded in the saturation scale of the proton and nucleus for simplicity. $\hat K_b$ is a normalization factor which takes account of information about a transverse area for overlap region between the projectile proton and the target nucleus. However, throughout this paper, we leave it an arbitrary constant, since we shall consider the ratio of the hadron multiplicity in rare events to that in minimum bias events.

For inclusive hadron production at finite transverse momentum, a light hadron FF ($D_h$) is involved with the gluon production cross-section, as usual. However, it is unclear whether fragmentation function is applicable to low $p_\perp$ hadron production. Nevertheless, we shall take into account gluon fragmentation function because such a fragmenting process can play a significant role to provide us with reliable predictive power to describe data of charged hadron production. We shall go though this further below.

In our numerical computations, we employ $D_h(z)=6.05 z^{-0.714}(1-z)^{2.92}$ which corresponds to the NLO parametrization of the Kniehl-Kramer-Potter (KKP) FF for charged hadron production at $\mu=2$\;GeV~\cite{Kniehl:2000fe}. Now charged hadron multiplicity at pseudorapidity $\eta$ can be written as
\begin{align}
\frac{dN_{ch}}{d\eta}=\frac{\hat{K}_{ch}}{\sigma_\text{inel}}\int d^2\bm{p}_\perp \int\limits^1_{z_\text{min}}dz\frac{D_h(z)}{z^2} J_{y\rightarrow\eta}\frac{d\sigma_{g}}{d^2\bm{p}_{g\perp} dy}
\label{eq:Nch-dy}
\end{align}
where $J_{y\rightarrow\eta}=p_{g\perp}\cosh\eta/\sqrt{p_{g\perp}^2\cosh^2\eta+m_h^2}$ is the Jacobian for transforming the expression in $y$-space to that in $\eta$-space. We have assumed that $y=y_h=y_g$ and defined $\bm{p}_\perp\equiv z \bm{p}_{g\perp}$ for simplicity.
$\sigma_\text{inel}$ is an inelastic cross-section in $p+A$ collisions. We will put a cut off $p_\text{max}=10$\;GeV and $p_\text{min}=0.1$\;GeV in Eq.~\eqref{eq:Nch-dy} in our numerical calculations. $z_\text{min}$ is determined from the kinematical condition, $x_{1,2}\leq1$.
The rapidity in $d\sigma_{g}/d^2\bm{p}_{g\perp}dy$ is replaced with
\begin{align}
y=\frac{1}{2}\ln\left[\frac{\sqrt{m_h^2+p_{g\perp}^2 \cosh^2\eta}+p_{g\perp}\sinh\eta}{\sqrt{m_h^2+p_{g\perp}^2 \cosh^2\eta}- p_{g\perp}\sinh\eta}\right]
\label{eq:y-eta}
\end{align}
where we assumed that hadron's transverse momentum is strongly correlated with the gluon's transverse momentum $p_{g\perp}$ so that we use $p_{g\perp}$ in the Jacobian and Eq.~\eqref{eq:y-eta}. With regard to the mass scale of the charged hadron, we fix $m_h$ as 300\;MeV. One must keep in mind that the rapidity of the produced gluon is shifted by $\Delta y=0.465$ as $y\rightarrow y-\Delta y$ in Eq.~\eqref{eq:y-eta} to perform numerical calculations in $p+A$ collisions at the LHC.

\section{Small-$x$ evolution}
%\label{subsection:BK}
\label{section:appendixD}

The rapidity or energy dependence of the dipole amplitude, to leading accuracy in $N_c$, is given by the nonlinear Balitsky-Kovchegov (BK) equation~\cite{Balitsky:1995ub,Kovchegov:1996ty}:
\begin{align}
-\frac{dD_{Y,\bm{r}_\perp}}{dY}
= \int d^2 \bm{r}_{1\perp} \mathcal{K}(r_\perp, r_{1\perp})
\Big [  D_{Y,\bm{r}_\perp} - D_{Y,\bm{r}_{1\perp}}D_{Y,\bm{r}_{2\perp}}
\Big ],
\label{eq:BK}
\end{align}
where the running coupling evolution kernel in Balitsky's prescription~\cite{Balitsky:2006wa} is given by
\begin{align}
\mathcal{K}(r_\perp,r_{1\perp})=&
\frac{\alpha_s (r_\perp^2) N_c} {2\pi^2}\,
\Bigg [
\frac{1}{r_{1\perp}^2} \left ( \frac{\alpha_s(r_{1\perp}^2)}{\alpha_s(r_{2\perp}^2)}-1  \right )
+
\frac{r_\perp^2}{r_{1\perp}^2 r_{2\perp}^2}
\non
&
+
\frac{1}{r_{2\perp}^2} \left ( \frac{\alpha_s(r_{2\perp}^2)}{\alpha_s(r_{1\perp}^2)}-1  \right )
\Bigg ],
\label{eq:rcBK-kernel}
\end{align}
with $\bm{r}_\perp= \bm{r}_{1\perp}+ \bm{r}_{2\perp}$ being the size of the parent dipole size prior to one step in $Y$ evolution. The one loop coupling constant in coordinate space $\alpha_s(r_\perp^2)= 1/\left [\frac{9}{4\pi} \ln \left (\frac{4 C^2}{r_\perp^2\Lambda^{2}}+\hat a \right ) \right ]$ is employed to solve the rcBK equation. We can use the initial dipole amplitude at $x=x_0=0.01$ or $Y_0=\ln1/x_0$ to be of the form given by the McLerran-Venugopalan (MV) model~\cite{McLerran:1993ni,McLerran:1993ka}:
\begin{align}
D_{Y=Y_0,\bm{r}_\perp}=
\exp\left[-\frac{\left(r_\perp^2Q_{sp,0}^2\right)^\gamma}{4}\ln\left(\frac{1}{r_\perp\Lambda}+e\right)\right],
\label{eq:IC-rcBK}
\end{align}
where $\gamma$ is an anomalous dimension, $Q_{sp,0}$ is the initial saturation scale in the proton at $x=x_0$. The infrared cutoff $\hat a$ is chosen by freezing $\alpha_s(r \to \infty)\equiv\alpha_\text{fr}$. For the initial input parameters in the rcBK equation, we set $Q_{sp,0}^2=0.168\;{\rm GeV}^2$, $\gamma=1.119$, $C=2.47$, $\Lambda=0.241\;\rm{GeV}$, and $\alpha_\text{fr}=1.0$. These parameters in this initial condition are obtained from global data fitting at HERA-DIS and given in Ref.~\cite{Albacete:2012xq,rcbk}. For the target nucleus,  $Q_{sA,0}^2=cA^{1/3}Q_{sp,0}^2$ where $c\lesssim0.5$ for minimum bias events in $p+A$ collisions is obtained from fitting the New Muon Collaboration data on the nuclear structure functions $F_{2,A}(x,Q^2)$~\cite{Dusling:2009ni}. For the purpose of our discussion, we shall fix simply $Q_{sA,0}^2=2\,Q_{sp,0}^2$ for heavy nuclei such as Pb and Au in our numerical calculations. Indeed, several previous studies~\cite{Ma:2015sia,Fujii:2015lld,Fujii:2017rqa,Ma:2017rsu} adopting the smaller value of $Q_{sA,0}^2$ succeeded in describing nuclear modification factor of $J/\psi$ and $D$ meson at RHIC and the LHC.

\begin{figure*}
	\centering
	\includegraphics[width=0.475\linewidth]{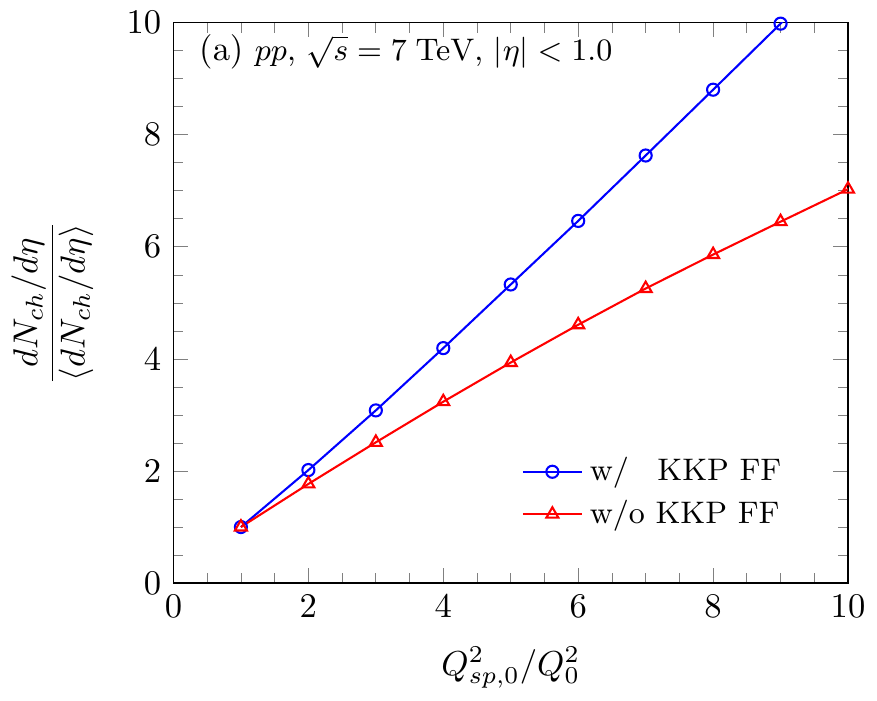}
	\includegraphics[width=0.475\linewidth]{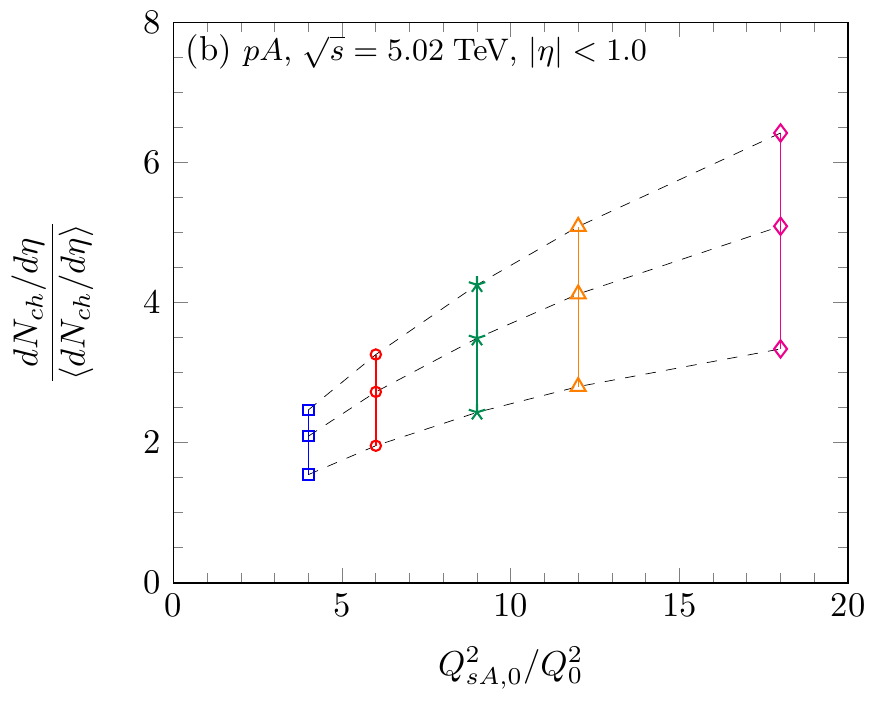}
	\caption{(Color online)
		(a): Relative multiplicity of charged hadrons  as a function of $Q_{sp,0}^2/Q_0^2$ with $Q_0^2=0.168\;\text{GeV}^2$ in $p+p$ collisions at the LHC. The same $Q_{sp,0}^2$ is applied to the projectile and the target. Solid (dashed) line is obtained with (without) use of the KKP FF. (b): Results in $p+A$ collisions are obtained by using the KKP FF and varying $Q_{sp,0}^2$ and $Q_{sA,0}^2$ independently.
	}
	\label{fig:Nch-Qs}
\end{figure*}

\begin{figure*}
	\centering
	\includegraphics[width=0.475\linewidth]{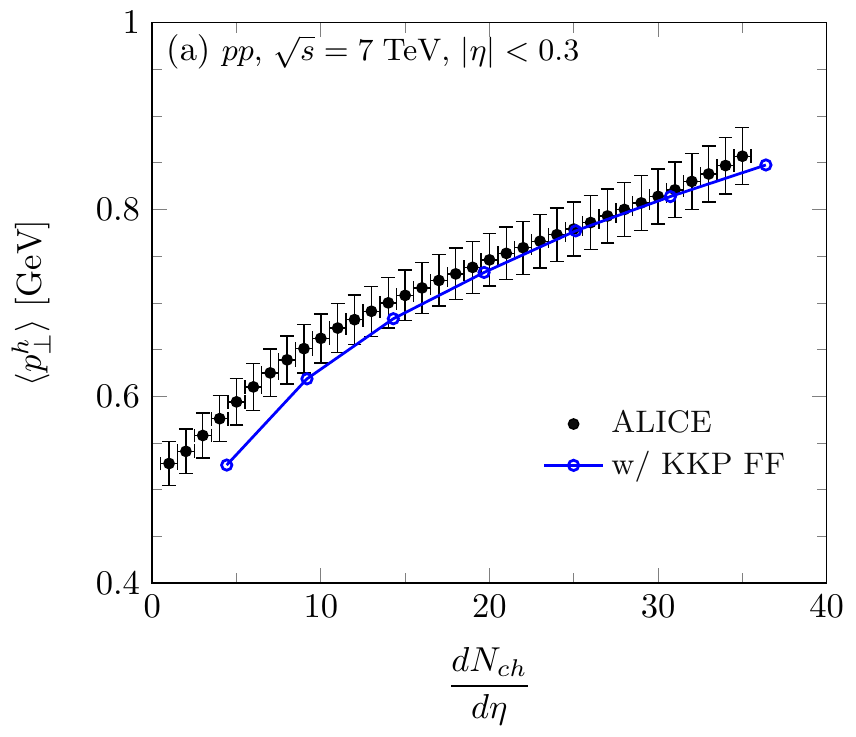}
	\includegraphics[width=0.475\linewidth]{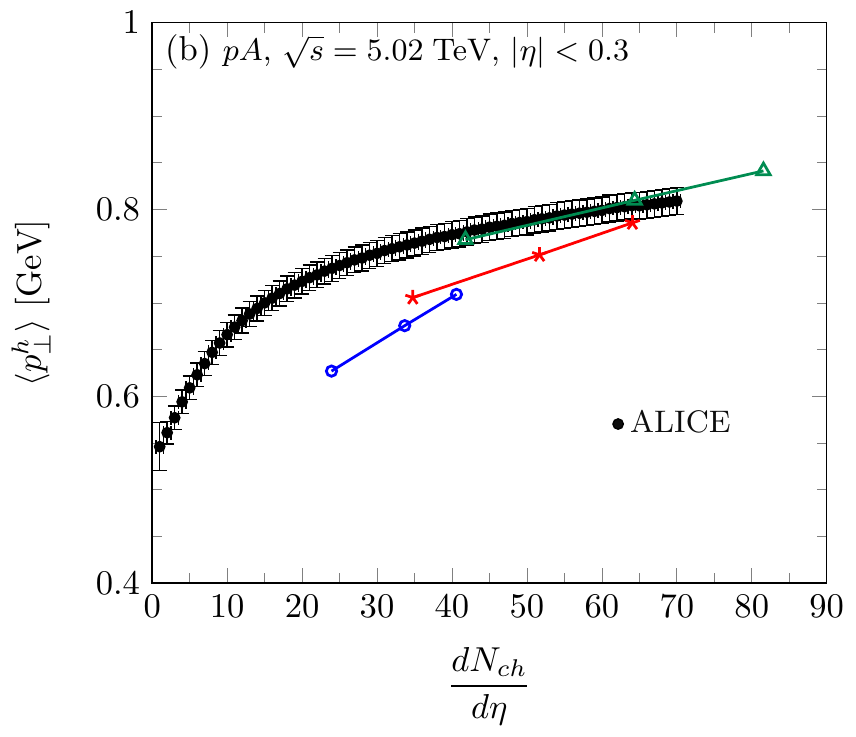}
	\caption{(Color online)
		Mean transverse momentum of produced hadron $h$ as a function of $dN_{ch}/d\eta$ in (a) $p+p$ and (b) $p+A$ collisions at the LHC in the mid rapidity region $|\eta_\text{lab}|<0.3$. Data are from Ref.~\cite{Abelev:2013bla}.}
	\label{fig:h-mean-Pt}
\end{figure*}

At large values of $x\geq x_0=0.01$, we need to extrapolate the parametrization of the dipole amplitude to these $x$ values. In Refs.~\cite{Ma:2014mri,Ma:2017rsu}, the adjoint dipole distribution in Eq.~\eqref{eq:proton-wavefn} at $x\geq x_0$ is determined to be $\mathcal{N}_{Y}^A(\bm{k}_\perp)\overset{x> x_0}{=}\;a(x)\mathcal{N}^A_{Y_0}(\bm{k}_\perp)$
where the coefficient $a(x)$ can be determined by matching the UGDF to collinear gluon distribution function. However, it is unclear whether the above matching procedure is applicable to high multiplicity events. In lieu, at large $x \geq x_0$, we adopt the simple extrapolation ansatz for \eqref{eq:proton-wavefn}~\cite{Gelis:2006tb}:
\begin{align}
\varphi_{p,y_p}(\bm{k}_\perp)=\varphi_{p,y_0}(\bm{k}_\perp)\left(\frac{1-x}{1-x_0}\right)^4 \left(\frac{x_0}{x}\right)^{0.15}.
\end{align}
We also apply the same procedure on the target side.

\section{Numerical results for inclusive hadron production}
%\label{subsection:inclusive-hadron-results}
\label{section:appendixE}

We first clarify our setup for numerical calculations in this paper. Assuming the CGC framework is yet applicable to $p+p$ collisions at collider energies, the only deference between $p+p$ collisions and $p+A$ collisions is the initial saturation scale for the target modulo the geometrical transverse size of the target. Regarding input parameters, we do not set $\hat{K}_b$, $\hat{K}_{ch}$, and $\sigma_\text{inel}$ to specific values here and leave these factors arbitrary in our numerical computations, since those parameters are irrelevant to the relative yield of $N_{ch}$. With regard to strong coupling constant $\alpha_s$ in Eqs.~\eqref{eq:xsection-kt-factorization-LN}\eqref{eq:xsection-kt-factorization-LN-CS}\eqref{eq:xsection-kt-factorization-LN-CO} and Eq.~(\ref{eq:gluon-kt}), we fix it as a constant value like $\alpha_s\sim0.2$ because all the differential cross-sections in this paper have been derived at leading order in $\alpha_s$.

Figure~\ref{fig:Nch-Qs} shows relative $dN_{ch}/d\eta$ in $p+p$ collisions at the LHC at mid rapidity by varying the initial saturation scale $Q_{sp,0}^2$. We take the saturation scales of the projectile proton and the target proton to be symmetrical; $Q_{sp_1,0}^2=Q_{sp_2,0}^2$. The averaged $N_{ch}$ is obtained by setting $Q_{sp_1,0}^2=Q_{sp_2,0}^2=Q_0^2$ with $Q_0^2=0.168\;\text{GeV}^2$. The solid line is the result obtained by using the KKP FF, while the dashed lines correspond to the result without using the KKP FF. It is clear that the relative $N_{ch}$ grows almost linearly as $Q_{sp,0}^2$ increases when the KKP FF is used.

The computation of the multiplicity in $p+A$ collisions is generally more complicated because it depends on the combination of the saturation scale of the projectile proton and that of the target nucleus. In Fig.~\ref{fig:Nch-Qs} (b), several combinations of $Q_{sp,0}^2$ and $Q_{sA,0}^2$ are depicted in different lines. We set the averaged $N_{ch}$ in $p+A$ collisions as the result with $Q_{sp,0}^2=Q_0^2$ and $Q_{sA,0}^2=2Q_0^2$.
In contrast to $p+p$ collisions, the relative $N_{ch}$ in $p+A$ collisions does not show a rapid growth with increasing $Q_{sp,0}^2$ and $Q_{sA,0}^2$, even if we employ the KKP FF.

The mean transverse momentum $\langle p_\perp\rangle$ of hadrons produced in high multiplicity events in $p+p$ and $p+A$ collisions is an important observable to check whether the CGC framework describes bulk data. The definition of $\langle p_\perp\rangle$ is given by
\begin{align}
\langle p_\perp\rangle=\frac{\int d^2\bm{p}_\perp p_\perp \frac{d\sigma}{d^2\bm{p}_\perp dy}}{\int d^2\bm{p}_\perp \frac{d\sigma}{d^2\bm{p}_\perp dy}}.
\end{align}
Figure~\ref{fig:h-mean-Pt} shows $N_{ch}$ dependence of $\langle p_\perp\rangle$ for single hadron production in $p+p$ and $p+A$ collisions at the LHC. We fix normalization of $dN_{ch}/d\eta$ in $p+p$ and $p+A$ collisions to fit minimum bias data respectively. Using the KKP FF, one can obtain a reasonable description of the data in $p+p$ collisions at the LHC. In $p+A$ collisions, numerical results with larger saturation scales for the projectile proton and the target nucleus show a nice agreement with data at the highest multiplicity. These comparisons clearly substantiate the robustness of the CGC framework in describing bulk data.

\section{Additional numerical results for heavy-flavor cross-sections}
%\label{section:HFresults}
\label{section:appendixF}

\begin{figure*}
	\centering
	\includegraphics[width=0.475\linewidth]{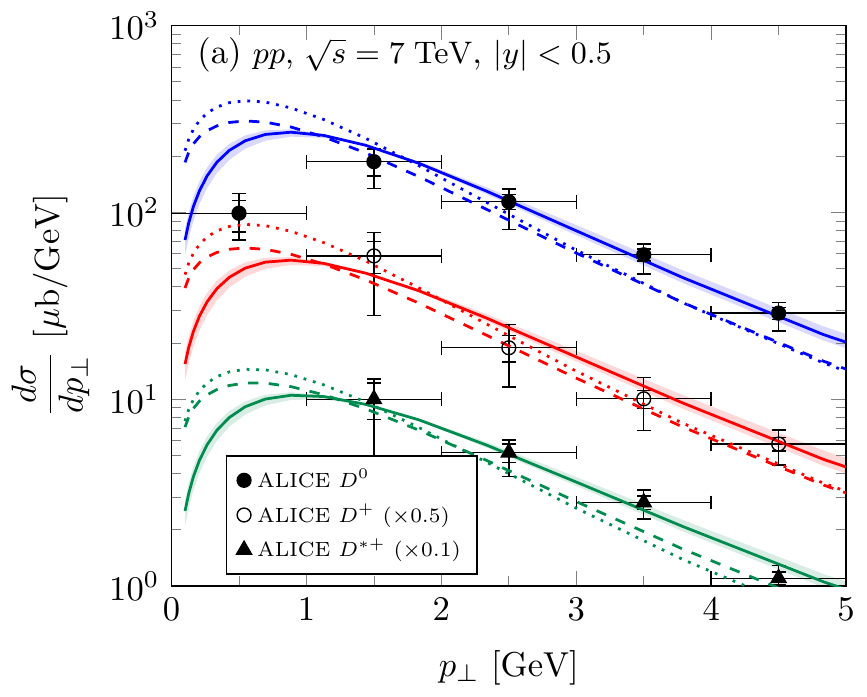}
	\includegraphics[width=0.475\linewidth]{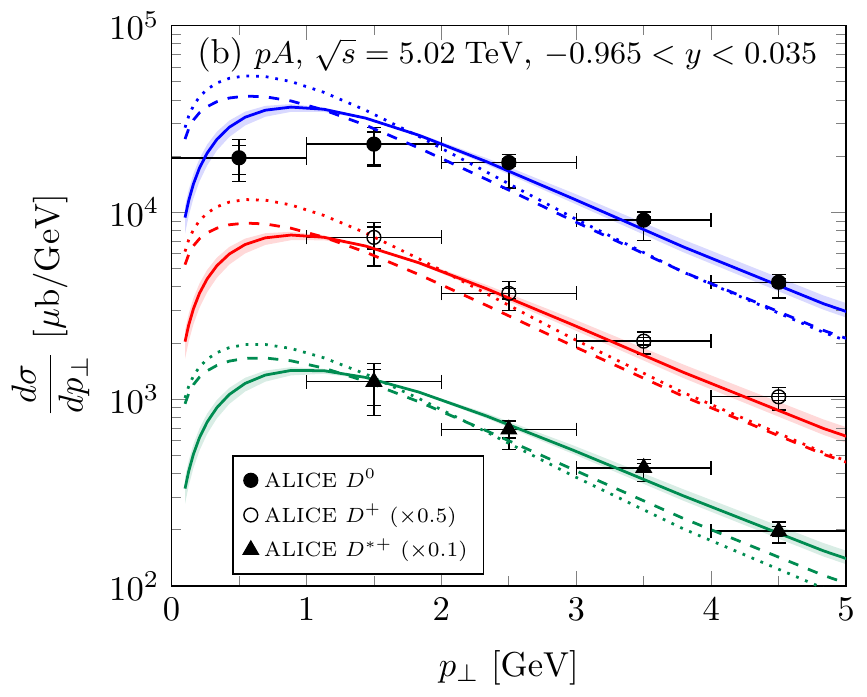}
	\caption{(Color online)
		Differential cross-sections for $D^0$ (blue), $D^+$ (red), $D^{\ast+}$ (green) production in (a) $p+p$ and (b) $p+A$ collisions at the LHC. The filled bands indicate uncertainties from the variations $r=0.06\textrm{--}0.135$ in the BCFY FFs (\ref{eq:BCFY-D0})-(\ref{eq:BCFY-Dstar}). The solid curves are obtained by setting $r=0.1$. Dashed (dotted) curves are obtained by using the BCFY FFs + DGLAP evolution (KKKS FFs) at $\mu=5$\;GeV. Data in $p+p$ collisions are taken from Refs.~\cite{Adam:2016ich,ALICE:2011aa}. Data in $p+A$ collisions are found in \cite{Adam:2016ich,Abelev:2014hha}.}
	\label{fig:D-pt-pp-pA}
\end{figure*}

\begin{figure*}
	\centering
	\includegraphics[width=0.475\linewidth]{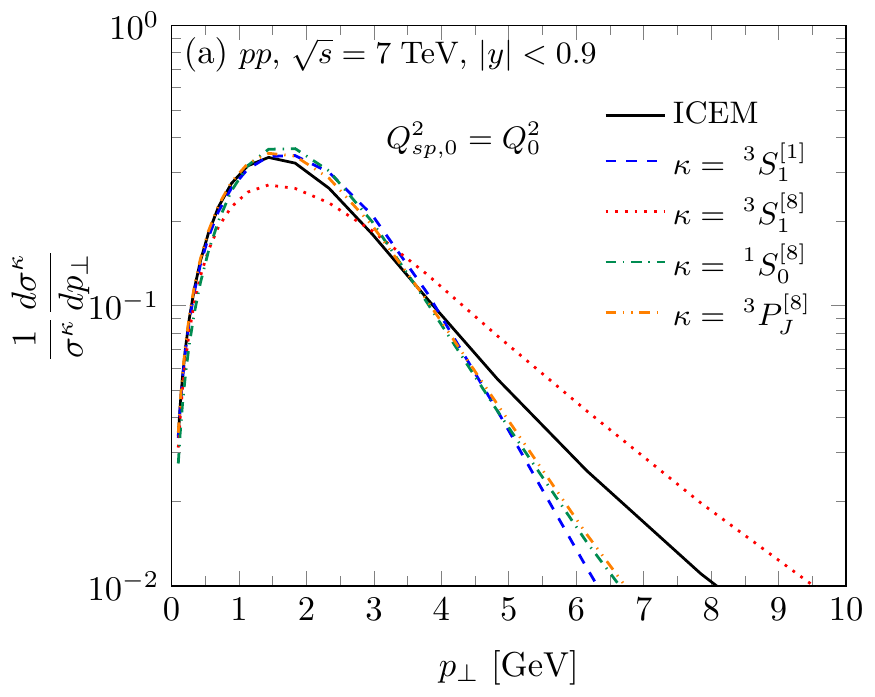}
	\includegraphics[width=0.475\linewidth]{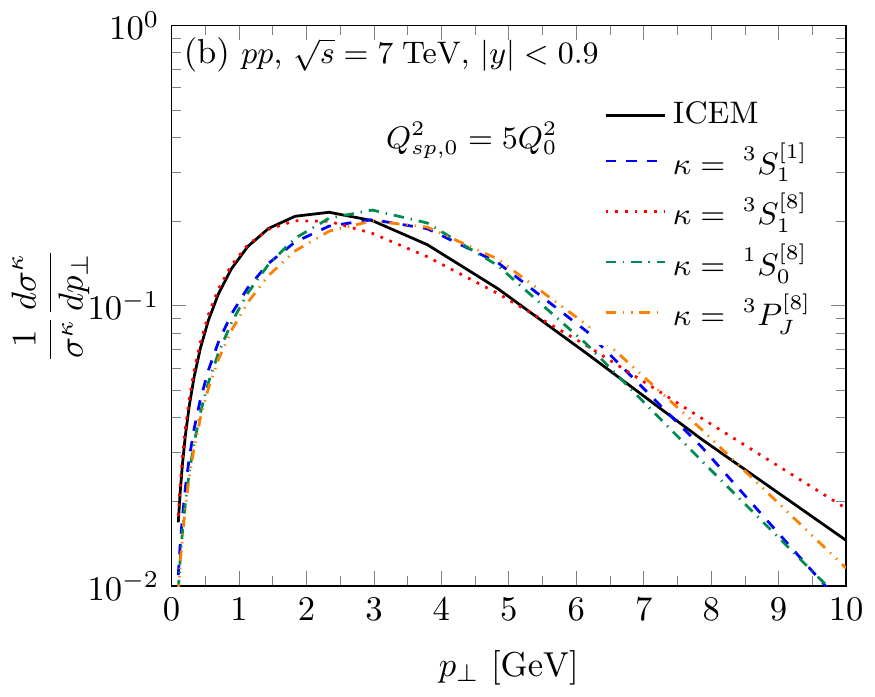}
	\caption{(Color online)
		Normalized differential cross-section of $c\bar c$ production for each $\kappa$ channel in minimum bias $p+p$ collisions at the LHC in the CGC+NRQCD framework along with the result in the CGC+ICEM model for (a) $Q_{sp,0}^2=Q_0^2$ and (b) $Q_{sp,0}^2=5Q_0^2$.
	}
	\label{fig:Jpsi-pp-pt}
\end{figure*}

\begin{figure*}
	\centering
	\includegraphics[width=0.475\linewidth]{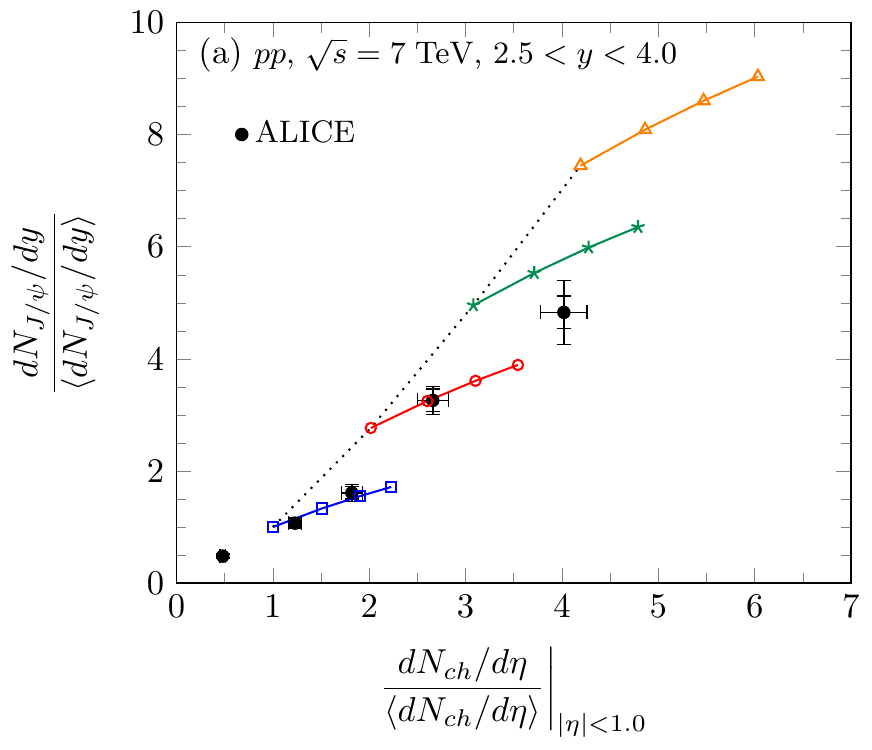}
	\includegraphics[width=0.475\linewidth]{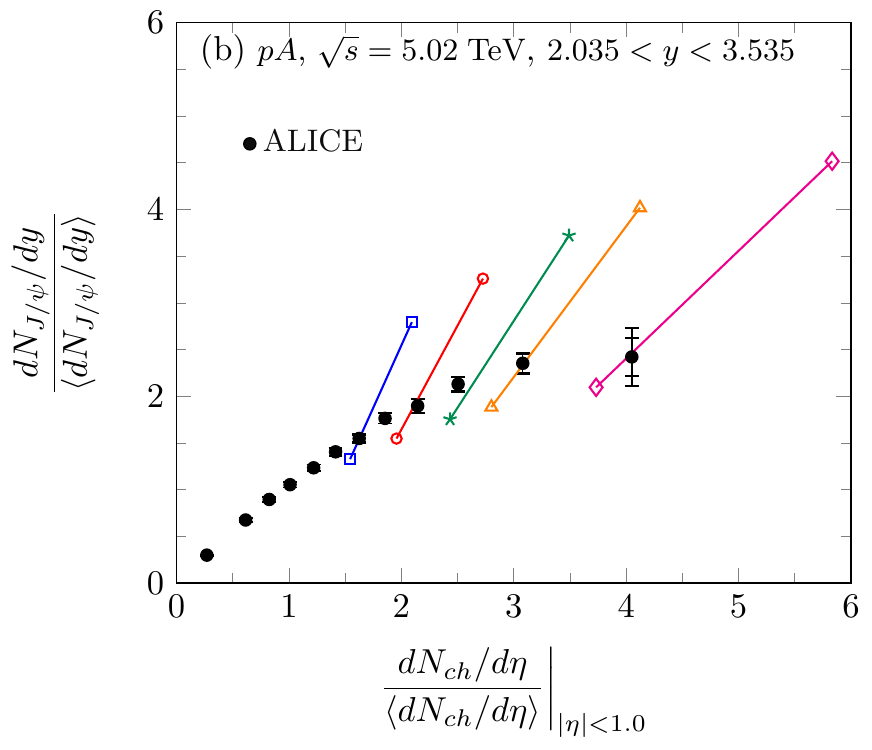}
	\caption{(Color online)
		(a): Results for forward $J/\psi$ production vs $N_{ch}$ in $p+p$ collisions at $\sqrt{s}=7\;$TeV  in the CGC+ICEM model. The blue points correspond to $Q_{sp1,0}^2=Q_0^2$ and $Q_{sp2,0}^2=1,2,3,4Q_0^2$. The red points correspond to $Q_{sp1,0}^2=2Q_0^2$ and $Q_{sp2,0}^2=2,3,4,5Q_0^2$. The green points correspond to $Q_{sp1,0}^2=3Q_0^2$ and $Q_{sp2,0}^2=3,4,5,6Q_0^2$. The orange points correspond to $Q_{sp1,0}^2=4Q_0^2$ and $Q_{sp2,0}^2=4,5,6,7Q_0^2$. Dotted line is obtained by taking $Q_{sp_1,0}=Q_{sp_2,0}$. (b): Results for forward $J/\psi$ production vs $N_{ch}$ in $p+A$ collisions at $\sqrt{s}=5.02\;$TeV/nucleon  in the CGC+ICEM model. The blue, red, green, orange, and magenta points all show model results for variations in the range $Q_{sp,0}^2 = \textrm{1--2} Q_0^2$ for $Q_{sA,0}^2=4,6,9,12, 24Q_0^2$ respectively. Data are from Ref.~\cite{Adamova:2017uhu}.}
	\label{fig:Jpsi-Nch-pA}
\end{figure*}

\begin{figure*}
	\centering
	\includegraphics[width=0.475\linewidth]{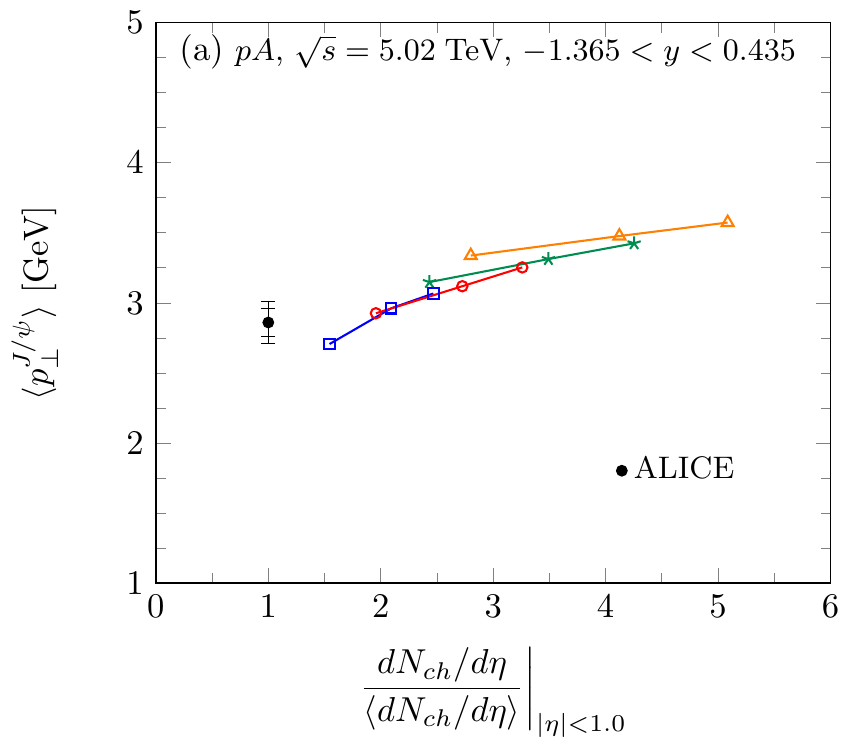}
	\includegraphics[width=0.475\linewidth]{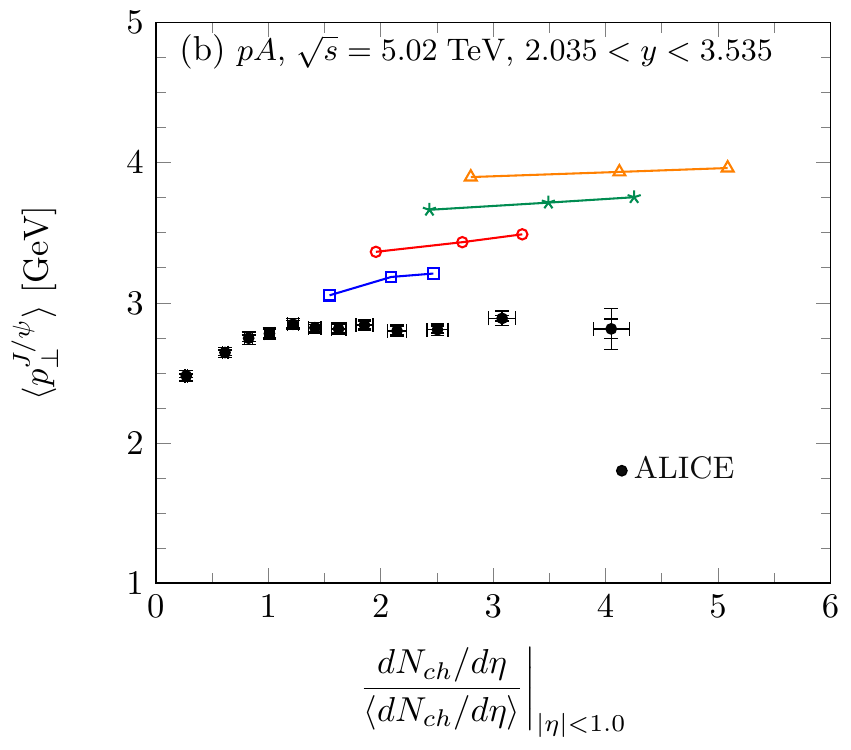}
	\caption{(Color online)
		Mean transverse momentum of $J/\psi$ as a function of $dN_{ch}/\langle dN_{ch}\rangle$ in $p+A$ collisions at the LHC at (a) $-1.365<y<0.435$ and (b) $2.035<y<3.535$. Data of $J/\psi$ production in minimum bias $p+A$ collisions are taken from Ref.~\cite{Adam:2015iga}. Data at forward rapidity are from \cite{Adamova:2017uhu}.}
	\label{fig:Jpsi-mean-Pt}
\end{figure*}

\begin{figure}
	\centering
	\includegraphics[width=0.95\linewidth]{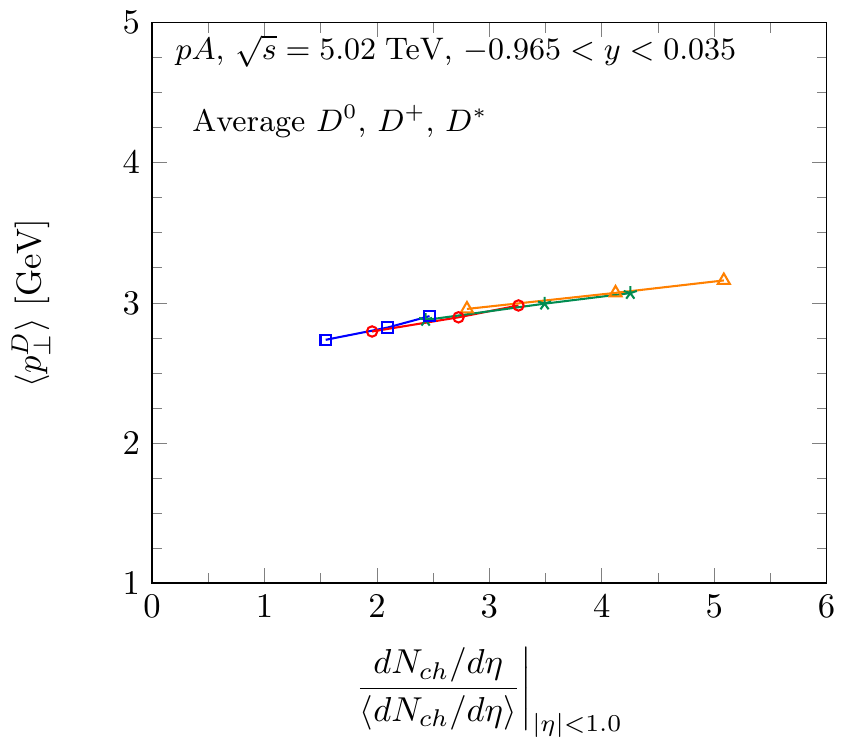}
	\caption{(Color online)
		Mean transverse momentum of average $D$ ($D^0$, $D^+$, $D^{\ast+}$) as a function of $dN_{ch}/\langle dN_{ch}\rangle$ in $p+A$ collisions at the LHC at $-0.965<y<0.035$. }
	\label{fig:D-mean-Pt}
\end{figure}

We will discuss here additional numerical results on cross-sections for $D$ and $J/\psi$ production. For the charm quark mass, we fixed $m_c=1.3\;$GeV in Eq.~\eqref{eq:xsection-kt-factorization-LN}, while $m_c=1.5\approx m_{J/\psi}/2$ is used in Eqs.~\eqref{eq:xsection-kt-factorization-LN-CS} and \eqref{eq:xsection-kt-factorization-LN-CO}. As noted in \cite{Ma:2014mri}, some of the dependence of on the quark masses in the short-distance cross-sections, is canceled out by the dependence of the LDMEs on quark mass.

In Fig.~\ref{fig:D-pt-pp-pA} (a), differential cross-sections for $D^0$, $D^+$, $D^\ast$ production in minimum bias $p+p$ collisions at the LHC are shown. As showed in Fig.~\ref{fig:dff-data-comparison}, the KKKS FFs agree quite well with $e^+e^-$ data relative to the BCFY FFs even after DGLAP evolution is taken into consideration. However, both these FF sets are in agreement with data on $D$ meson production in $p+p$ collisions for $p_\perp>1\;$GeV. Specifically, for the region in $p_\perp$ of  interest, from 1\;GeV to 4\;GeV, the BCFY curves and the KKKS curves are indistinguishable. Indeed, for the double ratio of minimum bias result to high multiplicity result, it makes little difference for our results. We, of course, anticipate that better data at high multiplicity can help us to confirm whether the tension with $e^+e^-$ data for the BCFY FFs is also seen in hadron-hadron collisions.

In Fig.~\ref{fig:D-pt-pp-pA} (a),
$K$-factor of 2.5 is required to describe data if we set the effective transverse area as $R_p=0.6\;$fm. However, a smaller value of $R_p$ can be also taken and is compatible with matching of unintegrated gluon distributions to gluon collinear PDFs at $x=0.01$~\cite{Ma:2014mri}. A smaller transverse area can therefore bring $50\%$ uncertainties to $K$ since higher order NLO effects cannot be distinguished from uncertainties in the transverse area. Indeed, this is a strong motivation for 
considering double ratios as we do, because the $K$-factor cancels out in the ratio.

In $p+A$ collisions, we determine the effective transverse area of the target nucleus $R_A$ by imposing that nuclear modification factor $R_{pA}=d\sigma_{pA}/(A d\sigma_{pp})$ for $c\bar c$ production should approach unity at asymptotically high $p_\perp$. This condition leads to $R_A=\sqrt{A/N^\gamma} R_p$ with $N=Q_{sA,0}^2/Q_{sp,0}^2$. Now the initial condition for $Q_{sA,0}^2=2Q_{sp,0}^2$ with $\gamma=1.119$ for the rcBK equation gives $R_A=9.79 R_p$. Using this value of $R_A$ with the same $K$-factor, Fig.~\ref{fig:D-pt-pp-pA} (b) shows a nice agreement with data in minimum bias $p+A$ collisions.

Figure~\ref {fig:Jpsi-pp-pt} (a) shows that for  minimum bias $Q_{sp,0}^2=Q_0^2$, the relative contributions of $d\sigma^\kappa/dp_\perp$ for  $\kappa = ^3S_1^{[1]}$, $^1S_0^{[8]}$, $^3P_J^{[8]}$ are similar to that of the ICEM at low $p_\perp$ and differs from the $^3S_1^{[8]}$. In contrast, the $p_\perp$ distribution of the latter is harder than the other channels at large $p_\perp$, a trend similar to that of the ICEM. This is understandable because high $p_\perp$ $J/\psi$ are likely to be produced via gluon fragmentation  with the quantum numbers of the $^3S_1^{[8]}$ channel. In contrast, Fig.~\ref {fig:Jpsi-pp-pt} (b) shows that for  rare $Q_{sp,0}^2 = 5\, Q_0^2$ configurations, the normalized $c\bar c$ differential cross-section for the $^3S_1^{[8]}$ channel is close to that of the ICEM over the entire $p_\perp$ range. The other channels are relatively harder at low $p_\perp$ and  softer at higher $p_\perp$.

We show in Fig.~\ref{fig:Jpsi-Nch-pA} comparisons of the ICEM with data on $N_{ch}$ dependence of $J/\psi$ production in $p+p$ and $p+A$ collisions at the LHC at forward rapidity. In contrast to mid rapidity, at forward rapidity, the symmetrical treatment; $Q_{sp_1,0}^2=Q_{sp_2,0}^2$ overshoots the data slightly in $p+p$ collisions. Data point at $dN_{ch}/\left<dN_{ch}\right>\sim4$ seems to favor the asymmetrical treatment; $Q_{sp_1,0}^2<Q_{sp_2,0}^2$. This is consistent with a naive expectation that a phase space for the gluon distribution of the projectile proton can shrink at forward rapidity ($x_1\sim{\cal O}(1)$) where a dilute-dense approximation is robust. One can find the similar trend for forward $J/\psi$ production in $p+A$ collisions.

Predictions for mean transverse momentum of $J/\psi$ production in $p+A$ collisions at the LHC are given in Fig.~\ref{fig:Jpsi-mean-Pt}. At mid rapidity, only the minimum bias data is available. The CGC prediction shows that $\langle p_\perp\rangle$ of $J/\psi$ depends on the change of the $Q_{sA,0}^2$ largely but does not change rapidly as $N_{ch}$ increases.  On the other hand, at forward rapidity, our numerical results overestimate $J/\psi$'s $\langle p_\perp\rangle$ at high $N_{ch}$. The comparable results for $\langle p_\perp \rangle$ of average $D$ ($D^0$, $D^+$, $D^\star$) production in $p+A$ collisions at mid-rapidity using the BCFY FFs with $r=0.1$ is shown in Fig.~\ref{fig:D-mean-Pt}, showing a relatively flat dependence on event activity compared to the $J/\psi$.

%%%%%%%%%%%%%%%%%%%%%%%%%%%%%%%%%%%%%%%%%%%%%%%%references
%\newpage
\bibliographystyle{apsrev4-1}
\bibliography{BibTexData}

\begin{thebibliography}{85}%
\makeatletter
\providecommand \@ifxundefined [1]{%
 \@ifx{#1\undefined}
}%
\providecommand \@ifnum [1]{%
 \ifnum #1\expandafter \@firstoftwo
 \else \expandafter \@secondoftwo
 \fi
}%
\providecommand \@ifx [1]{%
 \ifx #1\expandafter \@firstoftwo
 \else \expandafter \@secondoftwo
 \fi
}%
\providecommand \natexlab [1]{#1}%
\providecommand \enquote  [1]{``#1''}%
\providecommand \bibnamefont  [1]{#1}%
\providecommand \bibfnamefont [1]{#1}%
\providecommand \citenamefont [1]{#1}%
\providecommand \href@noop [0]{\@secondoftwo}%
\providecommand \href [0]{\begingroup \@sanitize@url \@href}%
\providecommand \@href[1]{\@@startlink{#1}\@@href}%
\providecommand \@@href[1]{\endgroup#1\@@endlink}%
\providecommand \@sanitize@url [0]{\catcode `\\12\catcode `\$12\catcode
  `\&12\catcode `\#12\catcode `\^12\catcode `\_12\catcode `\%12\relax}%
\providecommand \@@startlink[1]{}%
\providecommand \@@endlink[0]{}%
\providecommand \url  [0]{\begingroup\@sanitize@url \@url }%
\providecommand \@url [1]{\endgroup\@href {#1}{\urlprefix }}%
\providecommand \urlprefix  [0]{URL }%
\providecommand \Eprint [0]{\href }%
\providecommand \doibase [0]{http://dx.doi.org/}%
\providecommand \selectlanguage [0]{\@gobble}%
\providecommand \bibinfo  [0]{\@secondoftwo}%
\providecommand \bibfield  [0]{\@secondoftwo}%
\providecommand \translation [1]{[#1]}%
\providecommand \BibitemOpen [0]{}%
\providecommand \bibitemStop [0]{}%
\providecommand \bibitemNoStop [0]{.\EOS\space}%
\providecommand \EOS [0]{\spacefactor3000\relax}%
\providecommand \BibitemShut  [1]{\csname bibitem#1\endcsname}%
\let\auto@bib@innerbib\@empty
%</preamble>
\bibitem [{\citenamefont {Dumitru}\ \emph {et~al.}(2011)\citenamefont
  {Dumitru}, \citenamefont {Dusling}, \citenamefont {Gelis}, \citenamefont
  {Jalilian-Marian}, \citenamefont {Lappi},\ and\ \citenamefont
  {Venugopalan}}]{Dumitru:2010iy}%
  \BibitemOpen
  \bibfield  {author} {\bibinfo {author} {\bibfnamefont {A.}~\bibnamefont
  {Dumitru}}, \bibinfo {author} {\bibfnamefont {K.}~\bibnamefont {Dusling}},
  \bibinfo {author} {\bibfnamefont {F.}~\bibnamefont {Gelis}}, \bibinfo
  {author} {\bibfnamefont {J.}~\bibnamefont {Jalilian-Marian}}, \bibinfo
  {author} {\bibfnamefont {T.}~\bibnamefont {Lappi}}, \ and\ \bibinfo {author}
  {\bibfnamefont {R.}~\bibnamefont {Venugopalan}},\ }\href {\doibase
  10.1016/j.physletb.2011.01.024} {\bibfield  {journal} {\bibinfo  {journal}
  {Phys. Lett.}\ }\textbf {\bibinfo {volume} {B697}},\ \bibinfo {pages} {21}
  (\bibinfo {year} {2011})},\ \Eprint {http://arxiv.org/abs/1009.5295}
  {arXiv:1009.5295 [hep-ph]} \BibitemShut {NoStop}%
%%CITATION = ARXIV:1009.5295;%%
\bibitem [{\citenamefont {Dusling}\ and\ \citenamefont
  {Venugopalan}(2012)}]{Dusling:2012iga}%
  \BibitemOpen
  \bibfield  {author} {\bibinfo {author} {\bibfnamefont {K.}~\bibnamefont
  {Dusling}}\ and\ \bibinfo {author} {\bibfnamefont {R.}~\bibnamefont
  {Venugopalan}},\ }\href {\doibase 10.1103/PhysRevLett.108.262001} {\bibfield
  {journal} {\bibinfo  {journal} {Phys. Rev. Lett.}\ }\textbf {\bibinfo
  {volume} {108}},\ \bibinfo {pages} {262001} (\bibinfo {year} {2012})},\
  \Eprint {http://arxiv.org/abs/1201.2658} {arXiv:1201.2658 [hep-ph]}
  \BibitemShut {NoStop}%
%%CITATION = ARXIV:1201.2658;%%
\bibitem [{\citenamefont {Bozek}(2012)}]{Bozek:2011if}%
  \BibitemOpen
  \bibfield  {author} {\bibinfo {author} {\bibfnamefont {P.}~\bibnamefont
  {Bozek}},\ }\href {\doibase 10.1103/PhysRevC.85.014911} {\bibfield  {journal}
  {\bibinfo  {journal} {Phys. Rev.}\ }\textbf {\bibinfo {volume} {C85}},\
  \bibinfo {pages} {014911} (\bibinfo {year} {2012})},\ \Eprint
  {http://arxiv.org/abs/1112.0915} {arXiv:1112.0915 [hep-ph]} \BibitemShut
  {NoStop}%
%%CITATION = ARXIV:1112.0915;%%
\bibitem [{\citenamefont {Bozek}\ and\ \citenamefont
  {Broniowski}(2013)}]{Bozek:2012gr}%
  \BibitemOpen
  \bibfield  {author} {\bibinfo {author} {\bibfnamefont {P.}~\bibnamefont
  {Bozek}}\ and\ \bibinfo {author} {\bibfnamefont {W.}~\bibnamefont
  {Broniowski}},\ }\href {\doibase 10.1016/j.physletb.2012.12.051} {\bibfield
  {journal} {\bibinfo  {journal} {Phys. Lett.}\ }\textbf {\bibinfo {volume}
  {B718}},\ \bibinfo {pages} {1557} (\bibinfo {year} {2013})},\ \Eprint
  {http://arxiv.org/abs/1211.0845} {arXiv:1211.0845 [nucl-th]} \BibitemShut
  {NoStop}%
%%CITATION = ARXIV:1211.0845;%%
\bibitem [{\citenamefont {Bodwin}\ \emph {et~al.}(1995)\citenamefont {Bodwin},
  \citenamefont {Braaten},\ and\ \citenamefont {Lepage}}]{Bodwin:1994jh}%
  \BibitemOpen
  \bibfield  {author} {\bibinfo {author} {\bibfnamefont {G.~T.}\ \bibnamefont
  {Bodwin}}, \bibinfo {author} {\bibfnamefont {E.}~\bibnamefont {Braaten}}, \
  and\ \bibinfo {author} {\bibfnamefont {G.~P.}\ \bibnamefont {Lepage}},\
  }\href {\doibase 10.1103/PhysRevD.55.5853, 10.1103/PhysRevD.51.1125}
  {\bibfield  {journal} {\bibinfo  {journal} {Phys.Rev.}\ }\textbf {\bibinfo
  {volume} {D51}},\ \bibinfo {pages} {1125} (\bibinfo {year} {1995})},\ \Eprint
  {http://arxiv.org/abs/hep-ph/9407339} {arXiv:hep-ph/9407339 [hep-ph]}
  \BibitemShut {NoStop}%
%%CITATION = HEP-PH/9407339;%%
\bibitem [{\citenamefont {Ma}\ \emph {et~al.}(2011)\citenamefont {Ma},
  \citenamefont {Wang},\ and\ \citenamefont {Chao}}]{Ma:2010yw}%
  \BibitemOpen
  \bibfield  {author} {\bibinfo {author} {\bibfnamefont {Y.-Q.}\ \bibnamefont
  {Ma}}, \bibinfo {author} {\bibfnamefont {K.}~\bibnamefont {Wang}}, \ and\
  \bibinfo {author} {\bibfnamefont {K.-T.}\ \bibnamefont {Chao}},\ }\href
  {\doibase 10.1103/PhysRevLett.106.042002} {\bibfield  {journal} {\bibinfo
  {journal} {Phys.Rev.Lett.}\ }\textbf {\bibinfo {volume} {106}},\ \bibinfo
  {pages} {042002} (\bibinfo {year} {2011})},\ \Eprint
  {http://arxiv.org/abs/1009.3655} {arXiv:1009.3655 [hep-ph]} \BibitemShut
  {NoStop}%
%%CITATION = ARXIV:1009.3655;%%
\bibitem [{\citenamefont {Butenschon}\ and\ \citenamefont
  {Kniehl}(2011)}]{Butenschoen:2010rq}%
  \BibitemOpen
  \bibfield  {author} {\bibinfo {author} {\bibfnamefont {M.}~\bibnamefont
  {Butenschon}}\ and\ \bibinfo {author} {\bibfnamefont {B.~A.}\ \bibnamefont
  {Kniehl}},\ }\href {\doibase 10.1103/PhysRevLett.106.022003} {\bibfield
  {journal} {\bibinfo  {journal} {Phys.Rev.Lett.}\ }\textbf {\bibinfo {volume}
  {106}},\ \bibinfo {pages} {022003} (\bibinfo {year} {2011})},\ \Eprint
  {http://arxiv.org/abs/1009.5662} {arXiv:1009.5662 [hep-ph]} \BibitemShut
  {NoStop}%
%%CITATION = ARXIV:1009.5662;%%
\bibitem [{\citenamefont {Zhang}\ \emph {et~al.}(2010)\citenamefont {Zhang},
  \citenamefont {Ma}, \citenamefont {Wang},\ and\ \citenamefont
  {Chao}}]{Zhang:2009ym}%
  \BibitemOpen
  \bibfield  {author} {\bibinfo {author} {\bibfnamefont {Y.-J.}\ \bibnamefont
  {Zhang}}, \bibinfo {author} {\bibfnamefont {Y.-Q.}\ \bibnamefont {Ma}},
  \bibinfo {author} {\bibfnamefont {K.}~\bibnamefont {Wang}}, \ and\ \bibinfo
  {author} {\bibfnamefont {K.-T.}\ \bibnamefont {Chao}},\ }\href {\doibase
  10.1103/PhysRevD.81.034015} {\bibfield  {journal} {\bibinfo  {journal}
  {Phys.Rev.}\ }\textbf {\bibinfo {volume} {D81}},\ \bibinfo {pages} {034015}
  (\bibinfo {year} {2010})},\ \Eprint {http://arxiv.org/abs/0911.2166}
  {arXiv:0911.2166 [hep-ph]} \BibitemShut {NoStop}%
%%CITATION = ARXIV:0911.2166;%%
\bibitem [{\citenamefont {Gribov}\ \emph {et~al.}(1983)\citenamefont {Gribov},
  \citenamefont {Levin},\ and\ \citenamefont {Ryskin}}]{Gribov:1984tu}%
  \BibitemOpen
  \bibfield  {author} {\bibinfo {author} {\bibfnamefont {L.}~\bibnamefont
  {Gribov}}, \bibinfo {author} {\bibfnamefont {E.}~\bibnamefont {Levin}}, \
  and\ \bibinfo {author} {\bibfnamefont {M.}~\bibnamefont {Ryskin}},\ }\href
  {\doibase 10.1016/0370-1573(83)90022-4} {\bibfield  {journal} {\bibinfo
  {journal} {Phys.Rept.}\ }\textbf {\bibinfo {volume} {100}},\ \bibinfo {pages}
  {1} (\bibinfo {year} {1983})}\BibitemShut {NoStop}%
%%CITATION = PRPLC,100,1;%%
\bibitem [{\citenamefont {Mueller}\ and\ \citenamefont
  {Qiu}(1986)}]{Mueller:1985wy}%
  \BibitemOpen
  \bibfield  {author} {\bibinfo {author} {\bibfnamefont {A.~H.}\ \bibnamefont
  {Mueller}}\ and\ \bibinfo {author} {\bibfnamefont {J.}~\bibnamefont {Qiu}},\
  }\href {\doibase 10.1016/0550-3213(86)90164-1} {\bibfield  {journal}
  {\bibinfo  {journal} {Nucl.Phys.}\ }\textbf {\bibinfo {volume} {B268}},\
  \bibinfo {pages} {427} (\bibinfo {year} {1986})}\BibitemShut {NoStop}%
%%CITATION = NUPHA,B268,427;%%
\bibitem [{\citenamefont {Gelis}\ \emph {et~al.}(2010)\citenamefont {Gelis},
  \citenamefont {Iancu}, \citenamefont {Jalilian-Marian},\ and\ \citenamefont
  {Venugopalan}}]{Gelis:2010nm}%
  \BibitemOpen
  \bibfield  {author} {\bibinfo {author} {\bibfnamefont {F.}~\bibnamefont
  {Gelis}}, \bibinfo {author} {\bibfnamefont {E.}~\bibnamefont {Iancu}},
  \bibinfo {author} {\bibfnamefont {J.}~\bibnamefont {Jalilian-Marian}}, \ and\
  \bibinfo {author} {\bibfnamefont {R.}~\bibnamefont {Venugopalan}},\ }\href
  {\doibase 10.1146/annurev.nucl.010909.083629} {\bibfield  {journal} {\bibinfo
   {journal} {Ann.Rev.Nucl.Part.Sci.}\ }\textbf {\bibinfo {volume} {60}},\
  \bibinfo {pages} {463} (\bibinfo {year} {2010})},\ \Eprint
  {http://arxiv.org/abs/1002.0333} {arXiv:1002.0333 [hep-ph]} \BibitemShut
  {NoStop}%
%%CITATION = ARXIV:1002.0333;%%
\bibitem [{\citenamefont {Kovchegov}\ and\ \citenamefont
  {Levin}(2012)}]{Kovchegov:2012mbw}%
  \BibitemOpen
  \bibfield  {author} {\bibinfo {author} {\bibfnamefont {Y.~V.}\ \bibnamefont
  {Kovchegov}}\ and\ \bibinfo {author} {\bibfnamefont {E.}~\bibnamefont
  {Levin}},\ }\href@noop {} {\emph {\bibinfo {title} {{Quantum chromodynamics
  at high energy}}}}\ (\bibinfo {year} {2012})\BibitemShut {NoStop}%
%%CITATION = INSPIRE-1217905;%%
\bibitem [{\citenamefont {Blaizot}(2017)}]{Blaizot:2016qgz}%
  \BibitemOpen
  \bibfield  {author} {\bibinfo {author} {\bibfnamefont {J.-P.}\ \bibnamefont
  {Blaizot}},\ }\href {\doibase 10.1088/1361-6633/aa5435} {\bibfield  {journal}
  {\bibinfo  {journal} {Rept. Prog. Phys.}\ }\textbf {\bibinfo {volume} {80}},\
  \bibinfo {pages} {032301} (\bibinfo {year} {2017})},\ \Eprint
  {http://arxiv.org/abs/1607.04448} {arXiv:1607.04448 [hep-ph]} \BibitemShut
  {NoStop}%
%%CITATION = ARXIV:1607.04448;%%
\bibitem [{\citenamefont {Gelis}\ and\ \citenamefont
  {Venugopalan}(2004)}]{Gelis:2003vh}%
  \BibitemOpen
  \bibfield  {author} {\bibinfo {author} {\bibfnamefont {F.}~\bibnamefont
  {Gelis}}\ and\ \bibinfo {author} {\bibfnamefont {R.}~\bibnamefont
  {Venugopalan}},\ }\href {\doibase 10.1103/PhysRevD.69.014019} {\bibfield
  {journal} {\bibinfo  {journal} {Phys.Rev.}\ }\textbf {\bibinfo {volume}
  {D69}},\ \bibinfo {pages} {014019} (\bibinfo {year} {2004})},\ \Eprint
  {http://arxiv.org/abs/hep-ph/0310090} {arXiv:hep-ph/0310090 [hep-ph]}
  \BibitemShut {NoStop}%
%%CITATION = HEP-PH/0310090;%%
\bibitem [{\citenamefont {Blaizot}\ \emph
  {et~al.}(2004{\natexlab{a}})\citenamefont {Blaizot}, \citenamefont {Gelis},\
  and\ \citenamefont {Venugopalan}}]{Blaizot:2004wv}%
  \BibitemOpen
  \bibfield  {author} {\bibinfo {author} {\bibfnamefont {J.~P.}\ \bibnamefont
  {Blaizot}}, \bibinfo {author} {\bibfnamefont {F.}~\bibnamefont {Gelis}}, \
  and\ \bibinfo {author} {\bibfnamefont {R.}~\bibnamefont {Venugopalan}},\
  }\href {\doibase 10.1016/j.nuclphysa.2004.07.006} {\bibfield  {journal}
  {\bibinfo  {journal} {Nucl.Phys.}\ }\textbf {\bibinfo {volume} {A743}},\
  \bibinfo {pages} {57} (\bibinfo {year} {2004}{\natexlab{a}})},\ \Eprint
  {http://arxiv.org/abs/hep-ph/0402257} {arXiv:hep-ph/0402257 [hep-ph]}
  \BibitemShut {NoStop}%
%%CITATION = HEP-PH/0402257;%%
\bibitem [{\citenamefont {Tuchin}(2004)}]{Tuchin:2004rb}%
  \BibitemOpen
  \bibfield  {author} {\bibinfo {author} {\bibfnamefont {K.}~\bibnamefont
  {Tuchin}},\ }\href {\doibase 10.1016/j.physletb.2004.04.057} {\bibfield
  {journal} {\bibinfo  {journal} {Phys.Lett.}\ }\textbf {\bibinfo {volume}
  {B593}},\ \bibinfo {pages} {66} (\bibinfo {year} {2004})},\ \Eprint
  {http://arxiv.org/abs/hep-ph/0401022} {arXiv:hep-ph/0401022 [hep-ph]}
  \BibitemShut {NoStop}%
%%CITATION = HEP-PH/0401022;%%
\bibitem [{\citenamefont {Fujii}\ \emph {et~al.}(2005)\citenamefont {Fujii},
  \citenamefont {Gelis},\ and\ \citenamefont {Venugopalan}}]{Fujii:2005vj}%
  \BibitemOpen
  \bibfield  {author} {\bibinfo {author} {\bibfnamefont {H.}~\bibnamefont
  {Fujii}}, \bibinfo {author} {\bibfnamefont {F.}~\bibnamefont {Gelis}}, \ and\
  \bibinfo {author} {\bibfnamefont {R.}~\bibnamefont {Venugopalan}},\ }\href
  {\doibase 10.1103/PhysRevLett.95.162002} {\bibfield  {journal} {\bibinfo
  {journal} {Phys.Rev.Lett.}\ }\textbf {\bibinfo {volume} {95}},\ \bibinfo
  {pages} {162002} (\bibinfo {year} {2005})},\ \Eprint
  {http://arxiv.org/abs/hep-ph/0504047} {arXiv:hep-ph/0504047 [hep-ph]}
  \BibitemShut {NoStop}%
%%CITATION = HEP-PH/0504047;%%
\bibitem [{\citenamefont {Fujii}\ \emph {et~al.}(2006)\citenamefont {Fujii},
  \citenamefont {Gelis},\ and\ \citenamefont {Venugopalan}}]{Fujii:2006ab}%
  \BibitemOpen
  \bibfield  {author} {\bibinfo {author} {\bibfnamefont {H.}~\bibnamefont
  {Fujii}}, \bibinfo {author} {\bibfnamefont {F.}~\bibnamefont {Gelis}}, \ and\
  \bibinfo {author} {\bibfnamefont {R.}~\bibnamefont {Venugopalan}},\ }\href
  {\doibase 10.1016/j.nuclphysa.2006.09.012} {\bibfield  {journal} {\bibinfo
  {journal} {Nucl.Phys.}\ }\textbf {\bibinfo {volume} {A780}},\ \bibinfo
  {pages} {146} (\bibinfo {year} {2006})},\ \Eprint
  {http://arxiv.org/abs/hep-ph/0603099} {arXiv:hep-ph/0603099 [hep-ph]}
  \BibitemShut {NoStop}%
%%CITATION = HEP-PH/0603099;%%
\bibitem [{\citenamefont {Dominguez}\ \emph {et~al.}(2012)\citenamefont
  {Dominguez}, \citenamefont {Kharzeev}, \citenamefont {Levin}, \citenamefont
  {Mueller},\ and\ \citenamefont {Tuchin}}]{Dominguez:2011cy}%
  \BibitemOpen
  \bibfield  {author} {\bibinfo {author} {\bibfnamefont {F.}~\bibnamefont
  {Dominguez}}, \bibinfo {author} {\bibfnamefont {D.}~\bibnamefont {Kharzeev}},
  \bibinfo {author} {\bibfnamefont {E.}~\bibnamefont {Levin}}, \bibinfo
  {author} {\bibfnamefont {A.}~\bibnamefont {Mueller}}, \ and\ \bibinfo
  {author} {\bibfnamefont {K.}~\bibnamefont {Tuchin}},\ }\href {\doibase
  10.1016/j.physletb.2012.02.068} {\bibfield  {journal} {\bibinfo  {journal}
  {Phys.Lett.}\ }\textbf {\bibinfo {volume} {B710}},\ \bibinfo {pages} {182}
  (\bibinfo {year} {2012})},\ \Eprint {http://arxiv.org/abs/1109.1250}
  {arXiv:1109.1250 [hep-ph]} \BibitemShut {NoStop}%
%%CITATION = ARXIV:1109.1250;%%
\bibitem [{\citenamefont {Kang}\ \emph {et~al.}(2014)\citenamefont {Kang},
  \citenamefont {Ma},\ and\ \citenamefont {Venugopalan}}]{Kang:2013hta}%
  \BibitemOpen
  \bibfield  {author} {\bibinfo {author} {\bibfnamefont {Z.-B.}\ \bibnamefont
  {Kang}}, \bibinfo {author} {\bibfnamefont {Y.-Q.}\ \bibnamefont {Ma}}, \ and\
  \bibinfo {author} {\bibfnamefont {R.}~\bibnamefont {Venugopalan}},\ }\href
  {\doibase 10.1007/JHEP01(2014)056} {\bibfield  {journal} {\bibinfo  {journal}
  {JHEP}\ }\textbf {\bibinfo {volume} {1401}},\ \bibinfo {pages} {056}
  (\bibinfo {year} {2014})},\ \Eprint {http://arxiv.org/abs/1309.7337}
  {arXiv:1309.7337 [hep-ph]} \BibitemShut {NoStop}%
%%CITATION = ARXIV:1309.7337;%%
\bibitem [{\citenamefont {Fujii}\ and\ \citenamefont
  {Watanabe}(2013{\natexlab{a}})}]{Fujii:2013gxa}%
  \BibitemOpen
  \bibfield  {author} {\bibinfo {author} {\bibfnamefont {H.}~\bibnamefont
  {Fujii}}\ and\ \bibinfo {author} {\bibfnamefont {K.}~\bibnamefont
  {Watanabe}},\ }\href {\doibase 10.1016/j.nuclphysa.2013.06.011} {\bibfield
  {journal} {\bibinfo  {journal} {Nucl.Phys.}\ }\textbf {\bibinfo {volume}
  {A915}},\ \bibinfo {pages} {1} (\bibinfo {year} {2013}{\natexlab{a}})},\
  \Eprint {http://arxiv.org/abs/1304.2221} {arXiv:1304.2221 [hep-ph]}
  \BibitemShut {NoStop}%
%%CITATION = ARXIV:1304.2221;%%
\bibitem [{\citenamefont {Ma}\ and\ \citenamefont
  {Venugopalan}(2014)}]{Ma:2014mri}%
  \BibitemOpen
  \bibfield  {author} {\bibinfo {author} {\bibfnamefont {Y.-Q.}\ \bibnamefont
  {Ma}}\ and\ \bibinfo {author} {\bibfnamefont {R.}~\bibnamefont
  {Venugopalan}},\ }\href {\doibase 10.1103/PhysRevLett.113.192301} {\bibfield
  {journal} {\bibinfo  {journal} {Phys.Rev.Lett.}\ }\textbf {\bibinfo {volume}
  {113}},\ \bibinfo {pages} {192301} (\bibinfo {year} {2014})},\ \Eprint
  {http://arxiv.org/abs/1408.4075} {arXiv:1408.4075 [hep-ph]} \BibitemShut
  {NoStop}%
%%CITATION = ARXIV:1408.4075;%%
\bibitem [{\citenamefont {Ma}\ \emph {et~al.}(2015)\citenamefont {Ma},
  \citenamefont {Venugopalan},\ and\ \citenamefont {Zhang}}]{Ma:2015sia}%
  \BibitemOpen
  \bibfield  {author} {\bibinfo {author} {\bibfnamefont {Y.-Q.}\ \bibnamefont
  {Ma}}, \bibinfo {author} {\bibfnamefont {R.}~\bibnamefont {Venugopalan}}, \
  and\ \bibinfo {author} {\bibfnamefont {H.-F.}\ \bibnamefont {Zhang}},\ }\href
  {\doibase 10.1103/PhysRevD.92.071901} {\bibfield  {journal} {\bibinfo
  {journal} {Phys. Rev.}\ }\textbf {\bibinfo {volume} {D92}},\ \bibinfo {pages}
  {071901} (\bibinfo {year} {2015})},\ \Eprint
  {http://arxiv.org/abs/1503.07772} {arXiv:1503.07772 [hep-ph]} \BibitemShut
  {NoStop}%
%%CITATION = ARXIV:1503.07772;%%
\bibitem [{\citenamefont {Watanabe}\ and\ \citenamefont
  {Xiao}(2015)}]{Watanabe:2015yca}%
  \BibitemOpen
  \bibfield  {author} {\bibinfo {author} {\bibfnamefont {K.}~\bibnamefont
  {Watanabe}}\ and\ \bibinfo {author} {\bibfnamefont {B.-W.}\ \bibnamefont
  {Xiao}},\ }\href {\doibase 10.1103/PhysRevD.92.111502} {\bibfield  {journal}
  {\bibinfo  {journal} {Phys. Rev.}\ }\textbf {\bibinfo {volume} {D92}},\
  \bibinfo {pages} {111502} (\bibinfo {year} {2015})},\ \Eprint
  {http://arxiv.org/abs/1507.06564} {arXiv:1507.06564 [hep-ph]} \BibitemShut
  {NoStop}%
%%CITATION = ARXIV:1507.06564;%%
\bibitem [{\citenamefont {Duclou$\acute{\textrm{e}}$}\ \emph
  {et~al.}(2015)\citenamefont {Duclou$\acute{\textrm{e}}$}, \citenamefont
  {Lappi},\ and\ \citenamefont
  {M$\ddot{\textrm{a}}$ntysaari}}]{Ducloue:2015gfa}%
  \BibitemOpen
  \bibfield  {author} {\bibinfo {author} {\bibfnamefont {B.}~\bibnamefont
  {Duclou$\acute{\textrm{e}}$}}, \bibinfo {author} {\bibfnamefont
  {T.}~\bibnamefont {Lappi}}, \ and\ \bibinfo {author} {\bibfnamefont
  {H.}~\bibnamefont {M$\ddot{\textrm{a}}$ntysaari}},\ }\href@noop {} {\
  (\bibinfo {year} {2015})},\ \Eprint {http://arxiv.org/abs/1503.02789}
  {arXiv:1503.02789 [hep-ph]} \BibitemShut {NoStop}%
%%CITATION = ARXIV:1503.02789;%%
\bibitem [{\citenamefont {Fujii}\ and\ \citenamefont
  {Watanabe}(2013{\natexlab{b}})}]{Fujii:2013yja}%
  \BibitemOpen
  \bibfield  {author} {\bibinfo {author} {\bibfnamefont {H.}~\bibnamefont
  {Fujii}}\ and\ \bibinfo {author} {\bibfnamefont {K.}~\bibnamefont
  {Watanabe}},\ }\href {\doibase 10.1016/j.nuclphysa.2013.10.006} {\bibfield
  {journal} {\bibinfo  {journal} {Nucl. Phys.}\ }\textbf {\bibinfo {volume}
  {A920}},\ \bibinfo {pages} {78} (\bibinfo {year} {2013}{\natexlab{b}})},\
  \Eprint {http://arxiv.org/abs/1308.1258} {arXiv:1308.1258 [hep-ph]}
  \BibitemShut {NoStop}%
%%CITATION = ARXIV:1308.1258;%%
\bibitem [{\citenamefont {Fujii}\ and\ \citenamefont
  {Watanabe}(2016)}]{Fujii:2015lld}%
  \BibitemOpen
  \bibfield  {author} {\bibinfo {author} {\bibfnamefont {H.}~\bibnamefont
  {Fujii}}\ and\ \bibinfo {author} {\bibfnamefont {K.}~\bibnamefont
  {Watanabe}},\ }\href {\doibase 10.1016/j.nuclphysa.2016.03.045} {\bibfield
  {journal} {\bibinfo  {journal} {Nucl. Phys.}\ }\textbf {\bibinfo {volume}
  {A951}},\ \bibinfo {pages} {45} (\bibinfo {year} {2016})},\ \Eprint
  {http://arxiv.org/abs/1511.07698} {arXiv:1511.07698 [hep-ph]} \BibitemShut
  {NoStop}%
%%CITATION = ARXIV:1511.07698;%%
\bibitem [{\citenamefont {Duclou$\acute{\textrm{e}}$}\ \emph
  {et~al.}(2016)\citenamefont {Duclou$\acute{\textrm{e}}$}, \citenamefont
  {Lappi},\ and\ \citenamefont
  {M$\ddot{\textrm{a}}$ntysaari}}]{Ducloue:2016pqr}%
  \BibitemOpen
  \bibfield  {author} {\bibinfo {author} {\bibfnamefont {B.}~\bibnamefont
  {Duclou$\acute{\textrm{e}}$}}, \bibinfo {author} {\bibfnamefont
  {T.}~\bibnamefont {Lappi}}, \ and\ \bibinfo {author} {\bibfnamefont
  {H.}~\bibnamefont {M$\ddot{\textrm{a}}$ntysaari}},\ }\href {\doibase
  10.1103/PhysRevD.94.074031} {\bibfield  {journal} {\bibinfo  {journal} {Phys.
  Rev.}\ }\textbf {\bibinfo {volume} {D94}},\ \bibinfo {pages} {074031}
  (\bibinfo {year} {2016})},\ \Eprint {http://arxiv.org/abs/1605.05680}
  {arXiv:1605.05680 [hep-ph]} \BibitemShut {NoStop}%
%%CITATION = ARXIV:1605.05680;%%
\bibitem [{\citenamefont {Fujii}\ and\ \citenamefont
  {Watanabe}()}]{Fujii:2017rqa}%
  \BibitemOpen
  \bibfield  {author} {\bibinfo {author} {\bibfnamefont {H.}~\bibnamefont
  {Fujii}}\ and\ \bibinfo {author} {\bibfnamefont {K.}~\bibnamefont
  {Watanabe}},\ }\href@noop {} {\ }\Eprint {http://arxiv.org/abs/1706.06728}
  {arXiv:1706.06728 [hep-ph]} \BibitemShut {NoStop}%
%%CITATION = ARXIV:1706.06728;%%
\bibitem [{\citenamefont {Ma}\ \emph {et~al.}(2018)\citenamefont {Ma},
  \citenamefont {Venugopalan}, \citenamefont {Watanabe},\ and\ \citenamefont
  {Zhang}}]{Ma:2017rsu}%
  \BibitemOpen
  \bibfield  {author} {\bibinfo {author} {\bibfnamefont {Y.-Q.}\ \bibnamefont
  {Ma}}, \bibinfo {author} {\bibfnamefont {R.}~\bibnamefont {Venugopalan}},
  \bibinfo {author} {\bibfnamefont {K.}~\bibnamefont {Watanabe}}, \ and\
  \bibinfo {author} {\bibfnamefont {H.-F.}\ \bibnamefont {Zhang}},\ }\href
  {\doibase 10.1103/PhysRevC.97.014909} {\bibfield  {journal} {\bibinfo
  {journal} {Phys. Rev.}\ }\textbf {\bibinfo {volume} {C97}},\ \bibinfo {pages}
  {014909} (\bibinfo {year} {2018})},\ \Eprint
  {http://arxiv.org/abs/1707.07266} {arXiv:1707.07266 [hep-ph]} \BibitemShut
  {NoStop}%
%%CITATION = ARXIV:1707.07266;%%
\bibitem [{\citenamefont {Tribedy}\ and\ \citenamefont
  {Venugopalan}(2011)}]{Tribedy:2010ab}%
  \BibitemOpen
  \bibfield  {author} {\bibinfo {author} {\bibfnamefont {P.}~\bibnamefont
  {Tribedy}}\ and\ \bibinfo {author} {\bibfnamefont {R.}~\bibnamefont
  {Venugopalan}},\ }\href {\doibase 10.1016/j.nuclphysa.2011.04.008,
  10.1016/j.nuclphysa.2010.12.006} {\bibfield  {journal} {\bibinfo  {journal}
  {Nucl. Phys.}\ }\textbf {\bibinfo {volume} {A850}},\ \bibinfo {pages} {136}
  (\bibinfo {year} {2011})},\ \bibinfo {note} {[Erratum: Nucl.
  Phys.A859,185(2011)]},\ \Eprint {http://arxiv.org/abs/1011.1895}
  {arXiv:1011.1895 [hep-ph]} \BibitemShut {NoStop}%
%%CITATION = ARXIV:1011.1895;%%
\bibitem [{\citenamefont {Dusling}\ and\ \citenamefont
  {Venugopalan}(2013)}]{Dusling:2013qoz}%
  \BibitemOpen
  \bibfield  {author} {\bibinfo {author} {\bibfnamefont {K.}~\bibnamefont
  {Dusling}}\ and\ \bibinfo {author} {\bibfnamefont {R.}~\bibnamefont
  {Venugopalan}},\ }\href {\doibase 10.1103/PhysRevD.87.094034} {\bibfield
  {journal} {\bibinfo  {journal} {Phys. Rev.}\ }\textbf {\bibinfo {volume}
  {D87}},\ \bibinfo {pages} {094034} (\bibinfo {year} {2013})},\ \Eprint
  {http://arxiv.org/abs/1302.7018} {arXiv:1302.7018 [hep-ph]} \BibitemShut
  {NoStop}%
%%CITATION = ARXIV:1302.7018;%%
\bibitem [{\citenamefont {Dusling}\ \emph {et~al.}(2016)\citenamefont
  {Dusling}, \citenamefont {Tribedy},\ and\ \citenamefont
  {Venugopalan}}]{Dusling:2015rja}%
  \BibitemOpen
  \bibfield  {author} {\bibinfo {author} {\bibfnamefont {K.}~\bibnamefont
  {Dusling}}, \bibinfo {author} {\bibfnamefont {P.}~\bibnamefont {Tribedy}}, \
  and\ \bibinfo {author} {\bibfnamefont {R.}~\bibnamefont {Venugopalan}},\
  }\href {\doibase 10.1103/PhysRevD.93.014034} {\bibfield  {journal} {\bibinfo
  {journal} {Phys. Rev.}\ }\textbf {\bibinfo {volume} {D93}},\ \bibinfo {pages}
  {014034} (\bibinfo {year} {2016})},\ \Eprint
  {http://arxiv.org/abs/1509.04410} {arXiv:1509.04410 [hep-ph]} \BibitemShut
  {NoStop}%
%%CITATION = ARXIV:1509.04410;%%
\bibitem [{\citenamefont {Schenke}\ \emph {et~al.}(2016)\citenamefont
  {Schenke}, \citenamefont {Schlichting}, \citenamefont {Tribedy},\ and\
  \citenamefont {Venugopalan}}]{Schenke:2016lrs}%
  \BibitemOpen
  \bibfield  {author} {\bibinfo {author} {\bibfnamefont {B.}~\bibnamefont
  {Schenke}}, \bibinfo {author} {\bibfnamefont {S.}~\bibnamefont
  {Schlichting}}, \bibinfo {author} {\bibfnamefont {P.}~\bibnamefont
  {Tribedy}}, \ and\ \bibinfo {author} {\bibfnamefont {R.}~\bibnamefont
  {Venugopalan}},\ }\href {\doibase 10.1103/PhysRevLett.117.162301} {\bibfield
  {journal} {\bibinfo  {journal} {Phys. Rev. Lett.}\ }\textbf {\bibinfo
  {volume} {117}},\ \bibinfo {pages} {162301} (\bibinfo {year} {2016})},\
  \Eprint {http://arxiv.org/abs/1607.02496} {arXiv:1607.02496 [hep-ph]}
  \BibitemShut {NoStop}%
%%CITATION = ARXIV:1607.02496;%%
\bibitem [{\citenamefont {Dumitru}\ and\ \citenamefont
  {Skokov}(2017)}]{Dumitru:2017cwt}%
  \BibitemOpen
  \bibfield  {author} {\bibinfo {author} {\bibfnamefont {A.}~\bibnamefont
  {Dumitru}}\ and\ \bibinfo {author} {\bibfnamefont {V.}~\bibnamefont
  {Skokov}},\ }\href {\doibase 10.1103/PhysRevD.96.056029} {\bibfield
  {journal} {\bibinfo  {journal} {Phys. Rev.}\ }\textbf {\bibinfo {volume}
  {D96}},\ \bibinfo {pages} {056029} (\bibinfo {year} {2017})},\ \Eprint
  {http://arxiv.org/abs/1704.05917} {arXiv:1704.05917 [hep-ph]} \BibitemShut
  {NoStop}%
%%CITATION = ARXIV:1704.05917;%%
\bibitem [{\citenamefont {Dumitru}\ and\ \citenamefont
  {Skokov}(2018)}]{Dumitru:2017ftq}%
  \BibitemOpen
  \bibfield  {author} {\bibinfo {author} {\bibfnamefont {A.}~\bibnamefont
  {Dumitru}}\ and\ \bibinfo {author} {\bibfnamefont {V.}~\bibnamefont
  {Skokov}},\ }\href {\doibase 10.1051/epjconf/201817203009} {\bibfield
  {journal} {\bibinfo  {journal} {EPJ Web Conf.}\ }\textbf {\bibinfo {volume}
  {172}},\ \bibinfo {pages} {03009} (\bibinfo {year} {2018})},\ \Eprint
  {http://arxiv.org/abs/1710.05041} {arXiv:1710.05041 [hep-ph]} \BibitemShut
  {NoStop}%
%%CITATION = ARXIV:1710.05041;%%
\bibitem [{\citenamefont {Dumitru}\ \emph {et~al.}(2018)\citenamefont
  {Dumitru}, \citenamefont {Kapilevich},\ and\ \citenamefont
  {Skokov}}]{Dumitru:2018iko}%
  \BibitemOpen
  \bibfield  {author} {\bibinfo {author} {\bibfnamefont {A.}~\bibnamefont
  {Dumitru}}, \bibinfo {author} {\bibfnamefont {G.}~\bibnamefont {Kapilevich}},
  \ and\ \bibinfo {author} {\bibfnamefont {V.}~\bibnamefont {Skokov}},\ }\href
  {\doibase 10.1016/j.nuclphysa.2018.03.012} {\bibfield  {journal} {\bibinfo
  {journal} {Nucl. Phys.}\ }\textbf {\bibinfo {volume} {A974}},\ \bibinfo
  {pages} {106} (\bibinfo {year} {2018})},\ \Eprint
  {http://arxiv.org/abs/1802.06111} {arXiv:1802.06111 [hep-ph]} \BibitemShut
  {NoStop}%
%%CITATION = ARXIV:1802.06111;%%
\bibitem [{\citenamefont {Mueller}\ \emph
  {et~al.}(2013{\natexlab{a}})\citenamefont {Mueller}, \citenamefont {Xiao},\
  and\ \citenamefont {Yuan}}]{Mueller:2012uf}%
  \BibitemOpen
  \bibfield  {author} {\bibinfo {author} {\bibfnamefont {A.~H.}\ \bibnamefont
  {Mueller}}, \bibinfo {author} {\bibfnamefont {B.-W.}\ \bibnamefont {Xiao}}, \
  and\ \bibinfo {author} {\bibfnamefont {F.}~\bibnamefont {Yuan}},\ }\href
  {\doibase 10.1103/PhysRevLett.110.082301} {\bibfield  {journal} {\bibinfo
  {journal} {Phys. Rev. Lett.}\ }\textbf {\bibinfo {volume} {110}},\ \bibinfo
  {pages} {082301} (\bibinfo {year} {2013}{\natexlab{a}})},\ \Eprint
  {http://arxiv.org/abs/1210.5792} {arXiv:1210.5792 [hep-ph]} \BibitemShut
  {NoStop}%
%%CITATION = ARXIV:1210.5792;%%
\bibitem [{\citenamefont {Mueller}\ \emph
  {et~al.}(2013{\natexlab{b}})\citenamefont {Mueller}, \citenamefont {Xiao},\
  and\ \citenamefont {Yuan}}]{Mueller:2013wwa}%
  \BibitemOpen
  \bibfield  {author} {\bibinfo {author} {\bibfnamefont {A.~H.}\ \bibnamefont
  {Mueller}}, \bibinfo {author} {\bibfnamefont {B.-W.}\ \bibnamefont {Xiao}}, \
  and\ \bibinfo {author} {\bibfnamefont {F.}~\bibnamefont {Yuan}},\ }\href
  {\doibase 10.1103/PhysRevD.88.114010} {\bibfield  {journal} {\bibinfo
  {journal} {Phys. Rev.}\ }\textbf {\bibinfo {volume} {D88}},\ \bibinfo {pages}
  {114010} (\bibinfo {year} {2013}{\natexlab{b}})},\ \Eprint
  {http://arxiv.org/abs/1308.2993} {arXiv:1308.2993 [hep-ph]} \BibitemShut
  {NoStop}%
%%CITATION = ARXIV:1308.2993;%%
\bibitem [{\citenamefont {Qiu}\ \emph {et~al.}(2014)\citenamefont {Qiu},
  \citenamefont {Sun}, \citenamefont {Xiao},\ and\ \citenamefont
  {Yuan}}]{Qiu:2013qka}%
  \BibitemOpen
  \bibfield  {author} {\bibinfo {author} {\bibfnamefont {J.-W.}\ \bibnamefont
  {Qiu}}, \bibinfo {author} {\bibfnamefont {P.}~\bibnamefont {Sun}}, \bibinfo
  {author} {\bibfnamefont {B.-W.}\ \bibnamefont {Xiao}}, \ and\ \bibinfo
  {author} {\bibfnamefont {F.}~\bibnamefont {Yuan}},\ }\href {\doibase
  10.1103/PhysRevD.89.034007} {\bibfield  {journal} {\bibinfo  {journal}
  {Phys.Rev.}\ }\textbf {\bibinfo {volume} {D89}},\ \bibinfo {pages} {034007}
  (\bibinfo {year} {2014})},\ \Eprint {http://arxiv.org/abs/1310.2230}
  {arXiv:1310.2230 [hep-ph]} \BibitemShut {NoStop}%
%%CITATION = ARXIV:1310.2230;%%
\bibitem [{\citenamefont {Qiu}\ and\ \citenamefont
  {Watanabe}(2017)}]{Qiu:2017xbx}%
  \BibitemOpen
  \bibfield  {author} {\bibinfo {author} {\bibfnamefont {J.-W.}\ \bibnamefont
  {Qiu}}\ and\ \bibinfo {author} {\bibfnamefont {K.}~\bibnamefont {Watanabe}},\
  }\href {\doibase 10.22323/1.308.0024} {\bibfield  {journal} {\bibinfo
  {journal} {PoS}\ }\textbf {\bibinfo {volume} {QCDEV2017}},\ \bibinfo {pages}
  {024} (\bibinfo {year} {2017})},\ \Eprint {http://arxiv.org/abs/1710.06928}
  {arXiv:1710.06928 [hep-ph]} \BibitemShut {NoStop}%
%%CITATION = ARXIV:1710.06928;%%
\bibitem [{\citenamefont {Adam}\ \emph
  {et~al.}(2015{\natexlab{a}})\citenamefont {Adam} \emph
  {et~al.}}]{Adam:2015ota}%
  \BibitemOpen
  \bibfield  {author} {\bibinfo {author} {\bibfnamefont {J.}~\bibnamefont
  {Adam}} \emph {et~al.} (\bibinfo {collaboration} {ALICE}),\ }\href {\doibase
  10.1007/JHEP09(2015)148} {\bibfield  {journal} {\bibinfo  {journal} {JHEP}\
  }\textbf {\bibinfo {volume} {09}},\ \bibinfo {pages} {148} (\bibinfo {year}
  {2015}{\natexlab{a}})},\ \Eprint {http://arxiv.org/abs/1505.00664}
  {arXiv:1505.00664 [nucl-ex]} \BibitemShut {NoStop}%
%%CITATION = ARXIV:1505.00664;%%
\bibitem [{\citenamefont {Adam}\ \emph
  {et~al.}(2016{\natexlab{a}})\citenamefont {Adam} \emph
  {et~al.}}]{Adam:2016mkz}%
  \BibitemOpen
  \bibfield  {author} {\bibinfo {author} {\bibfnamefont {J.}~\bibnamefont
  {Adam}} \emph {et~al.} (\bibinfo {collaboration} {ALICE}),\ }\href {\doibase
  10.1007/JHEP08(2016)078} {\bibfield  {journal} {\bibinfo  {journal} {JHEP}\
  }\textbf {\bibinfo {volume} {08}},\ \bibinfo {pages} {078} (\bibinfo {year}
  {2016}{\natexlab{a}})},\ \Eprint {http://arxiv.org/abs/1602.07240}
  {arXiv:1602.07240 [nucl-ex]} \BibitemShut {NoStop}%
%%CITATION = ARXIV:1602.07240;%%
\bibitem [{\citenamefont {Abelev}\ \emph
  {et~al.}(2012{\natexlab{a}})\citenamefont {Abelev} \emph
  {et~al.}}]{Abelev:2012rz}%
  \BibitemOpen
  \bibfield  {author} {\bibinfo {author} {\bibfnamefont {B.}~\bibnamefont
  {Abelev}} \emph {et~al.} (\bibinfo {collaboration} {ALICE}),\ }\href
  {\doibase 10.1016/j.physletb.2012.04.052} {\bibfield  {journal} {\bibinfo
  {journal} {Phys. Lett.}\ }\textbf {\bibinfo {volume} {B712}},\ \bibinfo
  {pages} {165} (\bibinfo {year} {2012}{\natexlab{a}})},\ \Eprint
  {http://arxiv.org/abs/1202.2816} {arXiv:1202.2816 [hep-ex]} \BibitemShut
  {NoStop}%
%%CITATION = ARXIV:1202.2816;%%
\bibitem [{\citenamefont {Weber}(2017)}]{Weber:2017hhm}%
  \BibitemOpen
  \bibfield  {author} {\bibinfo {author} {\bibfnamefont {S.~G.}\ \bibnamefont
  {Weber}} (\bibinfo {collaboration} {ALICE}),\ }\href {\doibase
  10.1016/j.nuclphysa.2017.06.054} {\bibfield  {journal} {\bibinfo  {journal}
  {Nucl. Phys.}\ }\textbf {\bibinfo {volume} {A967}},\ \bibinfo {pages} {333}
  (\bibinfo {year} {2017})},\ \Eprint {http://arxiv.org/abs/1704.04735}
  {arXiv:1704.04735 [hep-ex]} \BibitemShut {NoStop}%
%%CITATION = ARXIV:1704.04735;%%
\bibitem [{\citenamefont {Khatun}()}]{Khatun:2017yic}%
  \BibitemOpen
  \bibfield  {author} {\bibinfo {author} {\bibfnamefont {A.}~\bibnamefont
  {Khatun}} (\bibinfo {collaboration} {ALICE}),\ }\Eprint
  {http://arxiv.org/abs/1711.09865} {arXiv:1711.09865 [nucl-ex]} \BibitemShut
  {NoStop}%
%%CITATION = ARXIV:1711.09865;%%
\bibitem [{\citenamefont {Adamov\'a}\ \emph {et~al.}(2018)\citenamefont
  {Adamov\'a} \emph {et~al.}}]{Adamova:2017uhu}%
  \BibitemOpen
  \bibfield  {author} {\bibinfo {author} {\bibfnamefont {D.}~\bibnamefont
  {Adamov\'a}} \emph {et~al.} (\bibinfo {collaboration} {ALICE}),\ }\href
  {\doibase 10.1016/j.physletb.2017.11.008} {\bibfield  {journal} {\bibinfo
  {journal} {Phys. Lett.}\ }\textbf {\bibinfo {volume} {B776}},\ \bibinfo
  {pages} {91} (\bibinfo {year} {2018})},\ \Eprint
  {http://arxiv.org/abs/1704.00274} {arXiv:1704.00274 [nucl-ex]} \BibitemShut
  {NoStop}%
%%CITATION = ARXIV:1704.00274;%%
\bibitem [{\citenamefont {Ma}(2016)}]{Ma:2015xta}%
  \BibitemOpen
  \bibfield  {author} {\bibinfo {author} {\bibfnamefont {R.}~\bibnamefont {Ma}}
  (\bibinfo {collaboration} {STAR}),\ }\href {\doibase
  10.1016/j.nuclphysbps.2016.05.059} {\bibfield  {journal} {\bibinfo  {journal}
  {Nucl. Part. Phys. Proc.}\ }\textbf {\bibinfo {volume} {276-278}},\ \bibinfo
  {pages} {261} (\bibinfo {year} {2016})},\ \Eprint
  {http://arxiv.org/abs/1509.06440} {arXiv:1509.06440 [nucl-ex]} \BibitemShut
  {NoStop}%
%%CITATION = ARXIV:1509.06440;%%
\bibitem [{\citenamefont {Federi\v{c}ov\'a}(2017)}]{Federicova:2017bmd}%
  \BibitemOpen
  \bibfield  {author} {\bibinfo {author} {\bibfnamefont {P.}~\bibnamefont
  {Federi\v{c}ov\'a}} (\bibinfo {collaboration} {STAR}),\ }\href {\doibase
  10.1051/epjconf/201713801017} {\bibfield  {journal} {\bibinfo  {journal} {EPJ
  Web Conf.}\ }\textbf {\bibinfo {volume} {138}},\ \bibinfo {pages} {01017}
  (\bibinfo {year} {2017})}\BibitemShut {NoStop}%
%%CITATION = 00776,138,01017;%%
\bibitem [{\citenamefont {Ferreiro}\ and\ \citenamefont
  {Pajares}(2012)}]{Ferreiro:2012fb}%
  \BibitemOpen
  \bibfield  {author} {\bibinfo {author} {\bibfnamefont {E.~G.}\ \bibnamefont
  {Ferreiro}}\ and\ \bibinfo {author} {\bibfnamefont {C.}~\bibnamefont
  {Pajares}},\ }\href {\doibase 10.1103/PhysRevC.86.034903} {\bibfield
  {journal} {\bibinfo  {journal} {Phys. Rev.}\ }\textbf {\bibinfo {volume}
  {C86}},\ \bibinfo {pages} {034903} (\bibinfo {year} {2012})},\ \Eprint
  {http://arxiv.org/abs/1203.5936} {arXiv:1203.5936 [hep-ph]} \BibitemShut
  {NoStop}%
%%CITATION = ARXIV:1203.5936;%%
\bibitem [{\citenamefont {Ferreiro}\ and\ \citenamefont
  {Pajares}(2015)}]{Ferreiro:2015gea}%
  \BibitemOpen
  \bibfield  {author} {\bibinfo {author} {\bibfnamefont {E.~G.}\ \bibnamefont
  {Ferreiro}}\ and\ \bibinfo {author} {\bibfnamefont {C.}~\bibnamefont
  {Pajares}},\ }\href@noop {} {\  (\bibinfo {year} {2015})},\ \Eprint
  {http://arxiv.org/abs/1501.03381} {arXiv:1501.03381 [hep-ph]} \BibitemShut
  {NoStop}%
%%CITATION = ARXIV:1501.03381;%%
\bibitem [{\citenamefont {Kopeliovich}\ \emph {et~al.}(2013)\citenamefont
  {Kopeliovich}, \citenamefont {Pirner}, \citenamefont {Potashnikova},
  \citenamefont {Reygers},\ and\ \citenamefont
  {Schmidt}}]{Kopeliovich:2013yfa}%
  \BibitemOpen
  \bibfield  {author} {\bibinfo {author} {\bibfnamefont {B.}~\bibnamefont
  {Kopeliovich}}, \bibinfo {author} {\bibfnamefont {H.}~\bibnamefont {Pirner}},
  \bibinfo {author} {\bibfnamefont {I.}~\bibnamefont {Potashnikova}}, \bibinfo
  {author} {\bibfnamefont {K.}~\bibnamefont {Reygers}}, \ and\ \bibinfo
  {author} {\bibfnamefont {I.}~\bibnamefont {Schmidt}},\ }\href {\doibase
  10.1103/PhysRevD.88.116002} {\bibfield  {journal} {\bibinfo  {journal}
  {Phys.Rev.}\ }\textbf {\bibinfo {volume} {D88}},\ \bibinfo {pages} {116002}
  (\bibinfo {year} {2013})},\ \Eprint {http://arxiv.org/abs/1308.3638}
  {arXiv:1308.3638 [hep-ph]} \BibitemShut {NoStop}%
%%CITATION = ARXIV:1308.3638;%%
\bibitem [{\citenamefont {Thakur}\ \emph {et~al.}(2018)\citenamefont {Thakur},
  \citenamefont {De}, \citenamefont {Sahoo},\ and\ \citenamefont
  {Dansana}}]{Thakur:2017kpv}%
  \BibitemOpen
  \bibfield  {author} {\bibinfo {author} {\bibfnamefont {D.}~\bibnamefont
  {Thakur}}, \bibinfo {author} {\bibfnamefont {S.}~\bibnamefont {De}}, \bibinfo
  {author} {\bibfnamefont {R.}~\bibnamefont {Sahoo}}, \ and\ \bibinfo {author}
  {\bibfnamefont {S.}~\bibnamefont {Dansana}},\ }\href {\doibase
  10.1103/PhysRevD.97.094002} {\bibfield  {journal} {\bibinfo  {journal} {Phys.
  Rev.}\ }\textbf {\bibinfo {volume} {D97}},\ \bibinfo {pages} {094002}
  (\bibinfo {year} {2018})},\ \Eprint {http://arxiv.org/abs/1709.06358}
  {arXiv:1709.06358 [hep-ph]} \BibitemShut {NoStop}%
%%CITATION = ARXIV:1709.06358;%%
\bibitem [{\citenamefont {Werner}\ \emph {et~al.}(2014)\citenamefont {Werner},
  \citenamefont {Guiot}, \citenamefont {Karpenko},\ and\ \citenamefont
  {Pierog}}]{Werner:2013tya}%
  \BibitemOpen
  \bibfield  {author} {\bibinfo {author} {\bibfnamefont {K.}~\bibnamefont
  {Werner}}, \bibinfo {author} {\bibfnamefont {B.}~\bibnamefont {Guiot}},
  \bibinfo {author} {\bibfnamefont {I.}~\bibnamefont {Karpenko}}, \ and\
  \bibinfo {author} {\bibfnamefont {T.}~\bibnamefont {Pierog}},\ }\href
  {\doibase 10.1103/PhysRevC.89.064903} {\bibfield  {journal} {\bibinfo
  {journal} {Phys. Rev.}\ }\textbf {\bibinfo {volume} {C89}},\ \bibinfo {pages}
  {064903} (\bibinfo {year} {2014})},\ \Eprint {http://arxiv.org/abs/1312.1233}
  {arXiv:1312.1233 [nucl-th]} \BibitemShut {NoStop}%
%%CITATION = ARXIV:1312.1233;%%
\bibitem [{\citenamefont {Braaten}\ \emph {et~al.}(1995)\citenamefont
  {Braaten}, \citenamefont {Cheung}, \citenamefont {Fleming},\ and\
  \citenamefont {Yuan}}]{Braaten:1994bz}%
  \BibitemOpen
  \bibfield  {author} {\bibinfo {author} {\bibfnamefont {E.}~\bibnamefont
  {Braaten}}, \bibinfo {author} {\bibfnamefont {K.-m.}\ \bibnamefont {Cheung}},
  \bibinfo {author} {\bibfnamefont {S.}~\bibnamefont {Fleming}}, \ and\
  \bibinfo {author} {\bibfnamefont {T.~C.}\ \bibnamefont {Yuan}},\ }\href
  {\doibase 10.1103/PhysRevD.51.4819} {\bibfield  {journal} {\bibinfo
  {journal} {Phys. Rev.}\ }\textbf {\bibinfo {volume} {D51}},\ \bibinfo {pages}
  {4819} (\bibinfo {year} {1995})},\ \Eprint
  {http://arxiv.org/abs/hep-ph/9409316} {arXiv:hep-ph/9409316 [hep-ph]}
  \BibitemShut {NoStop}%
%%CITATION = HEP-PH/9409316;%%
\bibitem [{\citenamefont {Kneesch}\ \emph {et~al.}(2008)\citenamefont
  {Kneesch}, \citenamefont {Kniehl}, \citenamefont {Kramer},\ and\
  \citenamefont {Schienbein}}]{Kneesch:2007ey}%
  \BibitemOpen
  \bibfield  {author} {\bibinfo {author} {\bibfnamefont {T.}~\bibnamefont
  {Kneesch}}, \bibinfo {author} {\bibfnamefont {B.~A.}\ \bibnamefont {Kniehl}},
  \bibinfo {author} {\bibfnamefont {G.}~\bibnamefont {Kramer}}, \ and\ \bibinfo
  {author} {\bibfnamefont {I.}~\bibnamefont {Schienbein}},\ }\href {\doibase
  10.1016/j.nuclphysb.2008.02.015} {\bibfield  {journal} {\bibinfo  {journal}
  {Nucl. Phys.}\ }\textbf {\bibinfo {volume} {B799}},\ \bibinfo {pages} {34}
  (\bibinfo {year} {2008})},\ \Eprint {http://arxiv.org/abs/0712.0481}
  {arXiv:0712.0481 [hep-ph]} \BibitemShut {NoStop}%
%%CITATION = ARXIV:0712.0481;%%
\bibitem [{\citenamefont {Ma}\ and\ \citenamefont {Vogt}(2016)}]{Ma:2016exq}%
  \BibitemOpen
  \bibfield  {author} {\bibinfo {author} {\bibfnamefont {Y.-Q.}\ \bibnamefont
  {Ma}}\ and\ \bibinfo {author} {\bibfnamefont {R.}~\bibnamefont {Vogt}},\
  }\href {\doibase 10.1103/PhysRevD.94.114029} {\bibfield  {journal} {\bibinfo
  {journal} {Phys. Rev.}\ }\textbf {\bibinfo {volume} {D94}},\ \bibinfo {pages}
  {114029} (\bibinfo {year} {2016})},\ \Eprint
  {http://arxiv.org/abs/1609.06042} {arXiv:1609.06042 [hep-ph]} \BibitemShut
  {NoStop}%
%%CITATION = ARXIV:1609.06042;%%
\bibitem [{\citenamefont {Fritzsch}(1977)}]{Fritzsch:1977ay}%
  \BibitemOpen
  \bibfield  {author} {\bibinfo {author} {\bibfnamefont {H.}~\bibnamefont
  {Fritzsch}},\ }\href {\doibase 10.1016/0370-2693(77)90108-3} {\bibfield
  {journal} {\bibinfo  {journal} {Phys.Lett.}\ }\textbf {\bibinfo {volume}
  {B67}},\ \bibinfo {pages} {217} (\bibinfo {year} {1977})}\BibitemShut
  {NoStop}%
%%CITATION = PHLTA,B67,217;%%
\bibitem [{\citenamefont {Gluck}\ \emph {et~al.}(1978)\citenamefont {Gluck},
  \citenamefont {Owens},\ and\ \citenamefont {Reya}}]{Gluck:1977zm}%
  \BibitemOpen
  \bibfield  {author} {\bibinfo {author} {\bibfnamefont {M.}~\bibnamefont
  {Gluck}}, \bibinfo {author} {\bibfnamefont {J.~F.}\ \bibnamefont {Owens}}, \
  and\ \bibinfo {author} {\bibfnamefont {E.}~\bibnamefont {Reya}},\ }\href
  {\doibase 10.1103/PhysRevD.17.2324} {\bibfield  {journal} {\bibinfo
  {journal} {Phys. Rev.}\ }\textbf {\bibinfo {volume} {D17}},\ \bibinfo {pages}
  {2324} (\bibinfo {year} {1978})}\BibitemShut {NoStop}%
%%CITATION = PHRVA,D17,2324;%%
\bibitem [{\citenamefont {Barger}\ \emph {et~al.}(1980)\citenamefont {Barger},
  \citenamefont {Keung},\ and\ \citenamefont {Phillips}}]{Barger:1979js}%
  \BibitemOpen
  \bibfield  {author} {\bibinfo {author} {\bibfnamefont {V.~D.}\ \bibnamefont
  {Barger}}, \bibinfo {author} {\bibfnamefont {W.-Y.}\ \bibnamefont {Keung}}, \
  and\ \bibinfo {author} {\bibfnamefont {R.~J.~N.}\ \bibnamefont {Phillips}},\
  }\href {\doibase 10.1016/0370-2693(80)90444-X} {\bibfield  {journal}
  {\bibinfo  {journal} {Phys. Lett.}\ }\textbf {\bibinfo {volume} {91B}},\
  \bibinfo {pages} {253} (\bibinfo {year} {1980})}\BibitemShut {NoStop}%
%%CITATION = PHLTA,91B,253;%%
\bibitem [{\citenamefont {Tribedy}\ and\ \citenamefont
  {Venugopalan}(2012)}]{Tribedy:2011aa}%
  \BibitemOpen
  \bibfield  {author} {\bibinfo {author} {\bibfnamefont {P.}~\bibnamefont
  {Tribedy}}\ and\ \bibinfo {author} {\bibfnamefont {R.}~\bibnamefont
  {Venugopalan}},\ }\href {\doibase 10.1016/j.physletb.2012.02.047,
  10.1016/j.physletb.2012.12.004} {\bibfield  {journal} {\bibinfo  {journal}
  {Phys. Lett.}\ }\textbf {\bibinfo {volume} {B710}},\ \bibinfo {pages} {125}
  (\bibinfo {year} {2012})},\ \bibinfo {note} {[Erratum: Phys.
  Lett.B718,1154(2013)]},\ \Eprint {http://arxiv.org/abs/1112.2445}
  {arXiv:1112.2445 [hep-ph]} \BibitemShut {NoStop}%
%%CITATION = ARXIV:1112.2445;%%
\bibitem [{\citenamefont {Albacete}\ \emph {et~al.}(2013)\citenamefont
  {Albacete}, \citenamefont {Dumitru}, \citenamefont {Fujii},\ and\
  \citenamefont {Nara}}]{Albacete:2012xq}%
  \BibitemOpen
  \bibfield  {author} {\bibinfo {author} {\bibfnamefont {J.~L.}\ \bibnamefont
  {Albacete}}, \bibinfo {author} {\bibfnamefont {A.}~\bibnamefont {Dumitru}},
  \bibinfo {author} {\bibfnamefont {H.}~\bibnamefont {Fujii}}, \ and\ \bibinfo
  {author} {\bibfnamefont {Y.}~\bibnamefont {Nara}},\ }\href {\doibase
  10.1016/j.nuclphysa.2012.09.012} {\bibfield  {journal} {\bibinfo  {journal}
  {Nucl.Phys.}\ }\textbf {\bibinfo {volume} {A897}},\ \bibinfo {pages} {1}
  (\bibinfo {year} {2013})},\ \Eprint {http://arxiv.org/abs/1209.2001}
  {arXiv:1209.2001 [hep-ph]} \BibitemShut {NoStop}%
%%CITATION = ARXIV:1209.2001;%%
\bibitem [{\citenamefont {Tanji}\ and\ \citenamefont
  {Berges}(2018)}]{Tanji:2017xiw}%
  \BibitemOpen
  \bibfield  {author} {\bibinfo {author} {\bibfnamefont {N.}~\bibnamefont
  {Tanji}}\ and\ \bibinfo {author} {\bibfnamefont {J.}~\bibnamefont {Berges}},\
  }\href {\doibase 10.1103/PhysRevD.97.034013} {\bibfield  {journal} {\bibinfo
  {journal} {Phys. Rev.}\ }\textbf {\bibinfo {volume} {D97}},\ \bibinfo {pages}
  {034013} (\bibinfo {year} {2018})},\ \Eprint
  {http://arxiv.org/abs/1711.03445} {arXiv:1711.03445 [hep-ph]} \BibitemShut
  {NoStop}%
%%CITATION = ARXIV:1711.03445;%%
\bibitem [{\citenamefont {Acharya}\ \emph {et~al.}(2018)\citenamefont {Acharya}
  \emph {et~al.}}]{Acharya:2017tfn}%
  \BibitemOpen
  \bibfield  {author} {\bibinfo {author} {\bibfnamefont {S.}~\bibnamefont
  {Acharya}} \emph {et~al.} (\bibinfo {collaboration} {ALICE}),\ }\href
  {\doibase 10.1016/j.physletb.2018.02.039} {\bibfield  {journal} {\bibinfo
  {journal} {Phys. Lett.}\ }\textbf {\bibinfo {volume} {B780}},\ \bibinfo
  {pages} {7} (\bibinfo {year} {2018})},\ \Eprint
  {http://arxiv.org/abs/1709.06807} {arXiv:1709.06807 [nucl-ex]} \BibitemShut
  {NoStop}%
%%CITATION = ARXIV:1709.06807;%%
\bibitem [{\citenamefont {Artuso}\ \emph {et~al.}(2004)\citenamefont {Artuso}
  \emph {et~al.}}]{Artuso:2004pj}%
  \BibitemOpen
  \bibfield  {author} {\bibinfo {author} {\bibfnamefont {M.}~\bibnamefont
  {Artuso}} \emph {et~al.} (\bibinfo {collaboration} {CLEO}),\ }\href {\doibase
  10.1103/PhysRevD.70.112001} {\bibfield  {journal} {\bibinfo  {journal} {Phys.
  Rev.}\ }\textbf {\bibinfo {volume} {D70}},\ \bibinfo {pages} {112001}
  (\bibinfo {year} {2004})},\ \Eprint {http://arxiv.org/abs/hep-ex/0402040}
  {arXiv:hep-ex/0402040 [hep-ex]} \BibitemShut {NoStop}%
%%CITATION = HEP-EX/0402040;%%
\bibitem [{\citenamefont {Barate}\ \emph {et~al.}(2000)\citenamefont {Barate}
  \emph {et~al.}}]{Barate:1999bg}%
  \BibitemOpen
  \bibfield  {author} {\bibinfo {author} {\bibfnamefont {R.}~\bibnamefont
  {Barate}} \emph {et~al.} (\bibinfo {collaboration} {ALEPH}),\ }\href
  {\doibase 10.1007/s100520000421} {\bibfield  {journal} {\bibinfo  {journal}
  {Eur. Phys. J.}\ }\textbf {\bibinfo {volume} {C16}},\ \bibinfo {pages} {597}
  (\bibinfo {year} {2000})},\ \Eprint {http://arxiv.org/abs/hep-ex/9909032}
  {arXiv:hep-ex/9909032 [hep-ex]} \BibitemShut {NoStop}%
%%CITATION = HEP-EX/9909032;%%
\bibitem [{\citenamefont {Cacciari}\ \emph {et~al.}(2012)\citenamefont
  {Cacciari}, \citenamefont {Frixione}, \citenamefont {Houdeau}, \citenamefont
  {Mangano}, \citenamefont {Nason},\ and\ \citenamefont
  {Ridolfi}}]{Cacciari:2012ny}%
  \BibitemOpen
  \bibfield  {author} {\bibinfo {author} {\bibfnamefont {M.}~\bibnamefont
  {Cacciari}}, \bibinfo {author} {\bibfnamefont {S.}~\bibnamefont {Frixione}},
  \bibinfo {author} {\bibfnamefont {N.}~\bibnamefont {Houdeau}}, \bibinfo
  {author} {\bibfnamefont {M.~L.}\ \bibnamefont {Mangano}}, \bibinfo {author}
  {\bibfnamefont {P.}~\bibnamefont {Nason}}, \ and\ \bibinfo {author}
  {\bibfnamefont {G.}~\bibnamefont {Ridolfi}},\ }\href {\doibase
  10.1007/JHEP10(2012)137} {\bibfield  {journal} {\bibinfo  {journal} {JHEP}\
  }\textbf {\bibinfo {volume} {10}},\ \bibinfo {pages} {137} (\bibinfo {year}
  {2012})},\ \Eprint {http://arxiv.org/abs/1205.6344} {arXiv:1205.6344
  [hep-ph]} \BibitemShut {NoStop}%
%%CITATION = ARXIV:1205.6344;%%
\bibitem [{\citenamefont {Cacciari}\ and\ \citenamefont
  {Nason}(2003)}]{Cacciari:2003zu}%
  \BibitemOpen
  \bibfield  {author} {\bibinfo {author} {\bibfnamefont {M.}~\bibnamefont
  {Cacciari}}\ and\ \bibinfo {author} {\bibfnamefont {P.}~\bibnamefont
  {Nason}},\ }\href {\doibase 10.1088/1126-6708/2003/09/006} {\bibfield
  {journal} {\bibinfo  {journal} {JHEP}\ }\textbf {\bibinfo {volume} {09}},\
  \bibinfo {pages} {006} (\bibinfo {year} {2003})},\ \Eprint
  {http://arxiv.org/abs/hep-ph/0306212} {arXiv:hep-ph/0306212 [hep-ph]}
  \BibitemShut {NoStop}%
%%CITATION = HEP-PH/0306212;%%
\bibitem [{\citenamefont {Arleo}\ and\ \citenamefont
  {Guillet}()}]{ffgenerator}%
  \BibitemOpen
  \bibfield  {author} {\bibinfo {author} {\bibfnamefont {F.}~\bibnamefont
  {Arleo}}\ and\ \bibinfo {author} {\bibfnamefont {J.}~\bibnamefont
  {Guillet}},\ }\href@noop {} {\enquote {\bibinfo {title} {\textsc{FF}
  generator},}\ }\bibinfo {howpublished}
  {\url{http://lapth.cnrs.fr/ffgenerator/}}\BibitemShut {NoStop}%
\bibitem [{\citenamefont {Kovchegov}\ and\ \citenamefont
  {Tuchin}(2002)}]{Kovchegov:2001sc}%
  \BibitemOpen
  \bibfield  {author} {\bibinfo {author} {\bibfnamefont {Y.~V.}\ \bibnamefont
  {Kovchegov}}\ and\ \bibinfo {author} {\bibfnamefont {K.}~\bibnamefont
  {Tuchin}},\ }\href {\doibase 10.1103/PhysRevD.65.074026} {\bibfield
  {journal} {\bibinfo  {journal} {Phys. Rev.}\ }\textbf {\bibinfo {volume}
  {D65}},\ \bibinfo {pages} {074026} (\bibinfo {year} {2002})},\ \Eprint
  {http://arxiv.org/abs/hep-ph/0111362} {arXiv:hep-ph/0111362 [hep-ph]}
  \BibitemShut {NoStop}%
%%CITATION = HEP-PH/0111362;%%
\bibitem [{\citenamefont {Blaizot}\ \emph
  {et~al.}(2004{\natexlab{b}})\citenamefont {Blaizot}, \citenamefont {Gelis},\
  and\ \citenamefont {Venugopalan}}]{Blaizot:2004wu}%
  \BibitemOpen
  \bibfield  {author} {\bibinfo {author} {\bibfnamefont {J.~P.}\ \bibnamefont
  {Blaizot}}, \bibinfo {author} {\bibfnamefont {F.}~\bibnamefont {Gelis}}, \
  and\ \bibinfo {author} {\bibfnamefont {R.}~\bibnamefont {Venugopalan}},\
  }\href {\doibase 10.1016/j.nuclphysa.2004.07.005} {\bibfield  {journal}
  {\bibinfo  {journal} {Nucl.Phys.}\ }\textbf {\bibinfo {volume} {A743}},\
  \bibinfo {pages} {13} (\bibinfo {year} {2004}{\natexlab{b}})},\ \Eprint
  {http://arxiv.org/abs/hep-ph/0402256} {arXiv:hep-ph/0402256 [hep-ph]}
  \BibitemShut {NoStop}%
%%CITATION = HEP-PH/0402256;%%
\bibitem [{\citenamefont {Kniehl}\ \emph {et~al.}(2000)\citenamefont {Kniehl},
  \citenamefont {Kramer},\ and\ \citenamefont {Potter}}]{Kniehl:2000fe}%
  \BibitemOpen
  \bibfield  {author} {\bibinfo {author} {\bibfnamefont {B.~A.}\ \bibnamefont
  {Kniehl}}, \bibinfo {author} {\bibfnamefont {G.}~\bibnamefont {Kramer}}, \
  and\ \bibinfo {author} {\bibfnamefont {B.}~\bibnamefont {Potter}},\ }\href
  {\doibase 10.1016/S0550-3213(00)00303-5} {\bibfield  {journal} {\bibinfo
  {journal} {Nucl. Phys.}\ }\textbf {\bibinfo {volume} {B582}},\ \bibinfo
  {pages} {514} (\bibinfo {year} {2000})},\ \Eprint
  {http://arxiv.org/abs/hep-ph/0010289} {arXiv:hep-ph/0010289 [hep-ph]}
  \BibitemShut {NoStop}%
%%CITATION = HEP-PH/0010289;%%
\bibitem [{\citenamefont {Balitsky}(1996)}]{Balitsky:1995ub}%
  \BibitemOpen
  \bibfield  {author} {\bibinfo {author} {\bibfnamefont {I.}~\bibnamefont
  {Balitsky}},\ }\href {\doibase 10.1016/0550-3213(95)00638-9} {\bibfield
  {journal} {\bibinfo  {journal} {Nucl.Phys.}\ }\textbf {\bibinfo {volume}
  {B463}},\ \bibinfo {pages} {99} (\bibinfo {year} {1996})},\ \Eprint
  {http://arxiv.org/abs/hep-ph/9509348} {arXiv:hep-ph/9509348 [hep-ph]}
  \BibitemShut {NoStop}%
%%CITATION = HEP-PH/9509348;%%
\bibitem [{\citenamefont {Kovchegov}(1996)}]{Kovchegov:1996ty}%
  \BibitemOpen
  \bibfield  {author} {\bibinfo {author} {\bibfnamefont {Y.~V.}\ \bibnamefont
  {Kovchegov}},\ }\href {\doibase 10.1103/PhysRevD.54.5463} {\bibfield
  {journal} {\bibinfo  {journal} {Phys. Rev.}\ }\textbf {\bibinfo {volume}
  {D54}},\ \bibinfo {pages} {5463} (\bibinfo {year} {1996})},\ \Eprint
  {http://arxiv.org/abs/hep-ph/9605446} {arXiv:hep-ph/9605446 [hep-ph]}
  \BibitemShut {NoStop}%
%%CITATION = HEP-PH/9605446;%%
\bibitem [{\citenamefont {Balitsky}(2007)}]{Balitsky:2006wa}%
  \BibitemOpen
  \bibfield  {author} {\bibinfo {author} {\bibfnamefont {I.}~\bibnamefont
  {Balitsky}},\ }\href {\doibase 10.1103/PhysRevD.75.014001} {\bibfield
  {journal} {\bibinfo  {journal} {Phys. Rev.}\ }\textbf {\bibinfo {volume}
  {D75}},\ \bibinfo {pages} {014001} (\bibinfo {year} {2007})},\ \Eprint
  {http://arxiv.org/abs/hep-ph/0609105} {arXiv:hep-ph/0609105 [hep-ph]}
  \BibitemShut {NoStop}%
%%CITATION = HEP-PH/0609105;%%
\bibitem [{\citenamefont {McLerran}\ and\ \citenamefont
  {Venugopalan}(1994{\natexlab{a}})}]{McLerran:1993ni}%
  \BibitemOpen
  \bibfield  {author} {\bibinfo {author} {\bibfnamefont {L.~D.}\ \bibnamefont
  {McLerran}}\ and\ \bibinfo {author} {\bibfnamefont {R.}~\bibnamefont
  {Venugopalan}},\ }\href {\doibase 10.1103/PhysRevD.49.2233} {\bibfield
  {journal} {\bibinfo  {journal} {Phys.Rev.}\ }\textbf {\bibinfo {volume}
  {D49}},\ \bibinfo {pages} {2233} (\bibinfo {year} {1994}{\natexlab{a}})},\
  \Eprint {http://arxiv.org/abs/hep-ph/9309289} {arXiv:hep-ph/9309289 [hep-ph]}
  \BibitemShut {NoStop}%
%%CITATION = HEP-PH/9309289;%%
\bibitem [{\citenamefont {McLerran}\ and\ \citenamefont
  {Venugopalan}(1994{\natexlab{b}})}]{McLerran:1993ka}%
  \BibitemOpen
  \bibfield  {author} {\bibinfo {author} {\bibfnamefont {L.~D.}\ \bibnamefont
  {McLerran}}\ and\ \bibinfo {author} {\bibfnamefont {R.}~\bibnamefont
  {Venugopalan}},\ }\href {\doibase 10.1103/PhysRevD.49.3352} {\bibfield
  {journal} {\bibinfo  {journal} {Phys.Rev.}\ }\textbf {\bibinfo {volume}
  {D49}},\ \bibinfo {pages} {3352} (\bibinfo {year} {1994}{\natexlab{b}})},\
  \Eprint {http://arxiv.org/abs/hep-ph/9311205} {arXiv:hep-ph/9311205 [hep-ph]}
  \BibitemShut {NoStop}%
%%CITATION = HEP-PH/9311205;%%
\bibitem [{\citenamefont {Albacete}\ \emph {et~al.}()\citenamefont {Albacete},
  \citenamefont {Dumitru}, \citenamefont {Fujii},\ and\ \citenamefont
  {Nara}}]{rcbk}%
  \BibitemOpen
  \bibfield  {author} {\bibinfo {author} {\bibfnamefont {J.}~\bibnamefont
  {Albacete}}, \bibinfo {author} {\bibfnamefont {A.}~\bibnamefont {Dumitru}},
  \bibinfo {author} {\bibfnamefont {H.}~\bibnamefont {Fujii}}, \ and\ \bibinfo
  {author} {\bibfnamefont {Y.}~\bibnamefont {Nara}},\ }\href@noop {} {\enquote
  {\bibinfo {title} {mckt: version 1.30},}\ }\bibinfo {howpublished}
  {\url{http://faculty.baruch.cuny.edu/naturalscience/physics/dumitru/CGC_IC.html}}\BibitemShut
  {NoStop}%
\bibitem [{\citenamefont {Dusling}\ \emph {et~al.}(2010)\citenamefont
  {Dusling}, \citenamefont {Gelis}, \citenamefont {Lappi},\ and\ \citenamefont
  {Venugopalan}}]{Dusling:2009ni}%
  \BibitemOpen
  \bibfield  {author} {\bibinfo {author} {\bibfnamefont {K.}~\bibnamefont
  {Dusling}}, \bibinfo {author} {\bibfnamefont {F.}~\bibnamefont {Gelis}},
  \bibinfo {author} {\bibfnamefont {T.}~\bibnamefont {Lappi}}, \ and\ \bibinfo
  {author} {\bibfnamefont {R.}~\bibnamefont {Venugopalan}},\ }\href {\doibase
  10.1016/j.nuclphysa.2009.12.044} {\bibfield  {journal} {\bibinfo  {journal}
  {Nucl.Phys.}\ }\textbf {\bibinfo {volume} {A836}},\ \bibinfo {pages} {159}
  (\bibinfo {year} {2010})},\ \Eprint {http://arxiv.org/abs/0911.2720}
  {arXiv:0911.2720 [hep-ph]} \BibitemShut {NoStop}%
%%CITATION = ARXIV:0911.2720;%%
\bibitem [{\citenamefont {Abelev}\ \emph {et~al.}(2013)\citenamefont {Abelev}
  \emph {et~al.}}]{Abelev:2013bla}%
  \BibitemOpen
  \bibfield  {author} {\bibinfo {author} {\bibfnamefont {B.~B.}\ \bibnamefont
  {Abelev}} \emph {et~al.} (\bibinfo {collaboration} {ALICE}),\ }\href
  {\doibase 10.1016/j.physletb.2013.10.054} {\bibfield  {journal} {\bibinfo
  {journal} {Phys. Lett.}\ }\textbf {\bibinfo {volume} {B727}},\ \bibinfo
  {pages} {371} (\bibinfo {year} {2013})},\ \Eprint
  {http://arxiv.org/abs/1307.1094} {arXiv:1307.1094 [nucl-ex]} \BibitemShut
  {NoStop}%
%%CITATION = ARXIV:1307.1094;%%
\bibitem [{\citenamefont {Gelis}\ \emph {et~al.}(2006)\citenamefont {Gelis},
  \citenamefont {Stasto},\ and\ \citenamefont {Venugopalan}}]{Gelis:2006tb}%
  \BibitemOpen
  \bibfield  {author} {\bibinfo {author} {\bibfnamefont {F.}~\bibnamefont
  {Gelis}}, \bibinfo {author} {\bibfnamefont {A.~M.}\ \bibnamefont {Stasto}}, \
  and\ \bibinfo {author} {\bibfnamefont {R.}~\bibnamefont {Venugopalan}},\
  }\href {\doibase 10.1140/epjc/s10052-006-0020-x} {\bibfield  {journal}
  {\bibinfo  {journal} {Eur. Phys. J.}\ }\textbf {\bibinfo {volume} {C48}},\
  \bibinfo {pages} {489} (\bibinfo {year} {2006})},\ \Eprint
  {http://arxiv.org/abs/hep-ph/0605087} {arXiv:hep-ph/0605087 [hep-ph]}
  \BibitemShut {NoStop}%
%%CITATION = HEP-PH/0605087;%%
\bibitem [{\citenamefont {Adam}\ \emph
  {et~al.}(2016{\natexlab{b}})\citenamefont {Adam} \emph
  {et~al.}}]{Adam:2016ich}%
  \BibitemOpen
  \bibfield  {author} {\bibinfo {author} {\bibfnamefont {J.}~\bibnamefont
  {Adam}} \emph {et~al.} (\bibinfo {collaboration} {ALICE}),\ }\href {\doibase
  10.1103/PhysRevC.94.054908} {\bibfield  {journal} {\bibinfo  {journal} {Phys.
  Rev.}\ }\textbf {\bibinfo {volume} {C94}},\ \bibinfo {pages} {054908}
  (\bibinfo {year} {2016}{\natexlab{b}})},\ \Eprint
  {http://arxiv.org/abs/1605.07569} {arXiv:1605.07569 [nucl-ex]} \BibitemShut
  {NoStop}%
%%CITATION = ARXIV:1605.07569;%%
\bibitem [{\citenamefont {Abelev}\ \emph
  {et~al.}(2012{\natexlab{b}})\citenamefont {Abelev} \emph
  {et~al.}}]{ALICE:2011aa}%
  \BibitemOpen
  \bibfield  {author} {\bibinfo {author} {\bibfnamefont {B.}~\bibnamefont
  {Abelev}} \emph {et~al.} (\bibinfo {collaboration} {ALICE}),\ }\href
  {\doibase 10.1007/JHEP01(2012)128} {\bibfield  {journal} {\bibinfo  {journal}
  {JHEP}\ }\textbf {\bibinfo {volume} {01}},\ \bibinfo {pages} {128} (\bibinfo
  {year} {2012}{\natexlab{b}})},\ \Eprint {http://arxiv.org/abs/1111.1553}
  {arXiv:1111.1553 [hep-ex]} \BibitemShut {NoStop}%
%%CITATION = ARXIV:1111.1553;%%
\bibitem [{\citenamefont {Abelev}\ \emph {et~al.}(2014)\citenamefont {Abelev}
  \emph {et~al.}}]{Abelev:2014hha}%
  \BibitemOpen
  \bibfield  {author} {\bibinfo {author} {\bibfnamefont {B.~B.}\ \bibnamefont
  {Abelev}} \emph {et~al.} (\bibinfo {collaboration} {ALICE}),\ }\href
  {\doibase 10.1103/PhysRevLett.113.232301} {\bibfield  {journal} {\bibinfo
  {journal} {Phys. Rev. Lett.}\ }\textbf {\bibinfo {volume} {113}},\ \bibinfo
  {pages} {232301} (\bibinfo {year} {2014})},\ \Eprint
  {http://arxiv.org/abs/1405.3452} {arXiv:1405.3452 [nucl-ex]} \BibitemShut
  {NoStop}%
%%CITATION = ARXIV:1405.3452;%%
\bibitem [{\citenamefont {Adam}\ \emph
  {et~al.}(2015{\natexlab{b}})\citenamefont {Adam} \emph
  {et~al.}}]{Adam:2015iga}%
  \BibitemOpen
  \bibfield  {author} {\bibinfo {author} {\bibfnamefont {J.}~\bibnamefont
  {Adam}} \emph {et~al.} (\bibinfo {collaboration} {ALICE}),\ }\href {\doibase
  10.1007/JHEP06(2015)055} {\bibfield  {journal} {\bibinfo  {journal} {JHEP}\
  }\textbf {\bibinfo {volume} {06}},\ \bibinfo {pages} {055} (\bibinfo {year}
  {2015}{\natexlab{b}})},\ \Eprint {http://arxiv.org/abs/1503.07179}
  {arXiv:1503.07179 [nucl-ex]} \BibitemShut {NoStop}%
%%CITATION = ARXIV:1503.07179;%%
\end{thebibliography}%

\end{document}